\documentclass[aps,pra,twocolumn,superscriptaddress,nofootinbib]{revtex4}
\usepackage{amsmath}
\usepackage{amsfonts}
\usepackage{amssymb}
\usepackage{graphicx}
\usepackage{xcolor}
\usepackage{float}
\usepackage{hyperref}
\usepackage{lineno} % line numbers so the response letter can cite lines; remove for the final version

\newcommand{\ket}[1]{\ensuremath{\left|{#1}\right\rangle}}
\newcommand{\bra}[1]{\ensuremath{\left\langle{#1}\right |}}

\newcommand{\oper}[1]{\mathbf{\mathsf{#1}}}

\begin{document}

\title{Inline {inter-core} phase control in {multi-core} fiber with a commercial polarization controller}

\author{A. Contreras}
\affiliation{Departamento de F\'{\i}sica, Universidad de Concepci\'on, 160-C Concepci\'on, Chile}
\affiliation{ANID -- Millennium Science Initiative Program -- Millennium Institute for Research in Optics,
Universidad de Concepci\'on, 160-C Concepci\'on, Chile}
\author{A. Mu\~{n}oz}
\affiliation{Departamento de F\'{\i}sica, Facultad de Ciencias F\'isicas y Matem\'aticas, Universidad de Chile, Santiago, Chile}
\affiliation{ANID -- Millennium Science Initiative Program -- Millennium Institute for Research in Optics,
Universidad de Concepci\'on, 160-C Concepci\'on, Chile}
\author{G. dos Santos}
\affiliation{Departamento de F\'{\i}sica, Universidad de Concepci\'on, 160-C Concepci\'on, Chile}
\affiliation{ANID -- Millennium Science Initiative Program -- Millennium Institute for Research in Optics,
Universidad de Concepci\'on, 160-C Concepci\'on, Chile}
\author{G.~Ca\~{n}as}
\affiliation{Departamento de F\'isica, Universidad del B\'io-B\'io, Collao 1202, 5-C Concepci\'on, Chile}
\author{G. Saavedra}
\affiliation{Departamento de Ingenier\'{\i}a El\'ectrica, Universidad de Concepci\'on, 160-C Concepci\'on, Chile}
\author{C. Hermann-Avigliano}
\affiliation{Departamento de F\'{\i}sica, Facultad de Ciencias F\'isicas y Matem\'aticas, Universidad de Chile, Santiago, Chile}
\affiliation{ANID -- Millennium Science Initiative Program -- Millennium Institute for Research in Optics,
Universidad de Concepci\'on, 160-C Concepci\'on, Chile}
\author{E. S. G\'omez}
\affiliation{Departamento de F\'{\i}sica, Universidad de Concepci\'on, 160-C Concepci\'on, Chile}
\affiliation{ANID -- Millennium Science Initiative Program -- Millennium Institute for Research in Optics,
Universidad de Concepci\'on, 160-C Concepci\'on, Chile}
\author{G. Lima}
\affiliation{Departamento de F\'{\i}sica, Universidad de Concepci\'on, 160-C Concepci\'on, Chile}
\affiliation{ANID -- Millennium Science Initiative Program -- Millennium Institute for Research in Optics,
Universidad de Concepci\'on, 160-C Concepci\'on, Chile}
\author{S. P. Walborn}
\email{Corresponding author: swalborn@udec.cl}
\affiliation{Departamento de F\'{\i}sica, Universidad de Concepci\'on, 160-C Concepci\'on, Chile}
\affiliation{ANID -- Millennium Science Initiative Program -- Millennium Institute for Research in Optics,
Universidad de Concepci\'on, 160-C Concepci\'on, Chile}

\begin{abstract}
Inter-core phase control is a prerequisite for interferometry in multi-core fiber (MCF), but no compact inline element for this purpose is presently available, and existing approaches require fanout to single-mode fiber or free-space optics. We show that a commercial inline polarization controller, a low-loss and widely available component, can be operated as an inline inter-core phase shifter. Two-core interferometry on a jacketed
four-core patchcord gives a reversible inter-core phase of $0.95 \pm 0.02$~rad from a single
device, with an operating-point-independent lower bound of $0.38$~rad, driven through repeated
load-unload cycles at a rate set by the actuator. A diametrically compressed disk model is used to predict
the operating condition: with the input polarization aligned to a principal stress axis the device
imparts a relative phase between cores without rotating the polarization state of either, which we
verify by polarimetry on all four cores. The response is polarization dependent, falling to less than
$0.58$~rad for the orthogonal input. The result is a practical, inline route to
{inter-core} phase control and stabilization in all-MCF interferometers, requiring neither fanout to
single-mode fiber nor free-space optics.
\end{abstract}

%\pacs{42.25.Ja, 42.81.Gs, 42.79.Sz}
\maketitle

\section{Introduction}

Multi-core optical fibers (MCFs) have emerged as a powerful solution for the escalating demands of
both classical and quantum optical communications~\cite{xavier20}. Through space-division
multiplexing, each core acts as an independent optical channel, increasing aggregate bandwidth
while conserving physical footprint~\cite{richardson10,richardson13,Puttnam:21}. Beyond mere
parallelism, it is the {inter-core} phase stability between core modes that gives the platform its distinctive
capabilities for interferometric applications. In classical networks this {inter-core} phase underlies
interferometric core routing and fast core-selective switching~\cite{melo2025,BeamSteering}, key
building blocks for the SDM platforms developed to relieve the capacity limits of single-mode
fiber~\cite{Winzer:17,crosstalkdef}, and it is the basis of MCF interferometric sensors, in which
the cores act as the arms of a compact common-path interferometer so that an external perturbation
is transduced into an {inter-core} phase read out from the interference
spectrum~\cite{yao23,Zhao2020,Song2023,Suo2022,Cuando-Espitia2021}. In the quantum domain, MCFs
implement programmable multiport beam splitters and interferometers for high-dimensional quantum
information processing~\cite{carine21,taddei21,martinez22} and support the distribution of high-dimensional
quantum states and keys over deployed
fiber~\cite{lio20,gomez21,Wu2025,Xiong2018,santos2026highdimensionalquantumkeydistribution}. All of these applications suffer from the fact that in
fiber-based interferometers the phase drifts with temperature and mechanical stress and must be
actively controlled~\cite{carine20}. An inline element that imparts a controllable {inter-core}
phase, without removing light from the MCF, would therefore be a valuable tool across all of these
settings.

A distinct way in which mechanical effects couple to the guided light is through stress-induced
birefringence. A transverse stress makes the fiber locally birefringent and, in an MCF, imparts a
phase to each core that depends on its transverse position. This is the basis of a broad class of
MCF strain, load, bend, and shape sensors, in which an unknown external stress is reconstructed
from the induced birefringence and {inter-core}
phase~\cite{silva-lopez2004,silva-lopez2005,flockhart03,villatoro16,moore12,zhao2022advances}, and
of recent modeling of the intrinsic residual stress frozen into {multi-core}
preforms~\cite{ocampo25}. Mechanical perturbations are also known to couple the cores of an
MCF opto-mechanically, and guided acoustic modes launched in one core have been used to modulate
the light in a neighboring one~\cite{diamandi17}. In all of these cases the mechanical
perturbation is either unknown and to be reconstructed, as in sensing, or intrinsic to the fiber.
Here we invert the picture and ask whether a deliberately applied, externally controlled stress can
be used to impose a chosen {inter-core} phase. 

Ideally, the stress should be applied by a simple, inline, plug-and-play device.
Inline polarization controllers (IPCs), which apply a localized mechanical squeeze to the jacketed
fiber~\cite{Thorlabs_PolarizationController_2014,OptoSigma_PC_Web}, are a particularly good match: they impart a
well-defined stress geometry and act without disconnecting the fiber or removing light from the
system, requiring no fanout to single-mode fiber or free-space optics. These advantages suit
deployed and integrated systems, including all-MCF interferometers, multi-core quantum networks,
and inline sensing. 

In this paper we test a commercial IPC  as an {inter-core} phase controller on a jacketed fiber.
The way we operate the
device departs from its intended use. An IPC is normally adjusted by rotating the clamped fiber section
about the central axis while pressure is applied by the clamp, so that the birefringence axis is steered as the fiber is
twisted. Instead, we use the rotational degree of freedom to fix the load axis at a chosen angle
relative to the core lattice, which is a meaningful setting only for a {multi-core} fiber. Then the
pressure alone is varied.
As a theoretical guide, we adapt a diametrically-compressed-disk stress solution to the MCF geometry
and use it to describe both the intra-core birefringence and the {inter-core} phase imparted by an
IPC to a four-core fiber (Section \ref{sec:ipc}). Though this simplified model does not predict the magnitude of the {inter-core} phase, it yields a useful operating condition:
when the input polarization is aligned with a principal stress axis, the device acts as a pure
{inter-core} phase shifter, imparting a relative phase between cores without rotating the
polarization state of either. In Sec.~\ref{sec:exp} we present first a set of polarimetry experiments that qualitatively verify this on all
four cores independently. 
A second experiment measures the {inter-core} phase directly by two-core
interferometry. We obtained close to one radian of reversible, polarization-selectable phase from a
single off-the-shelf component. 

\section{Inline Polarization Controller and Stress-Induced Birefringence}
\label{sec:ipc}

\begin{figure}
\includegraphics[width=0.7\linewidth]{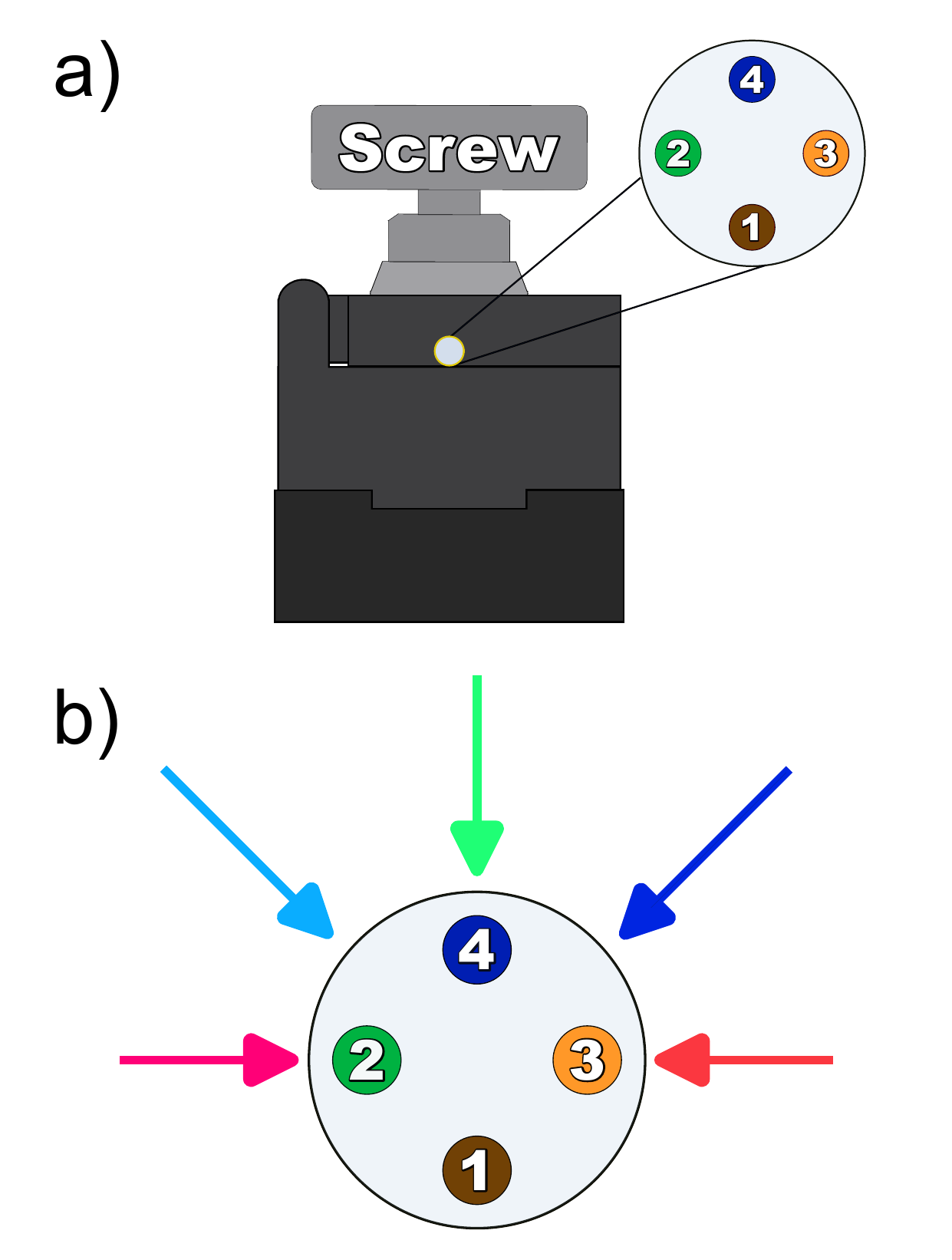}
\caption{a) Illustration of an inline polarization controller applied to a single-mode fiber and
MCF. b) Different pressure directions applied to the {multi-core} fiber. Red corresponds to a
horizontal force from the right (rh); blue to a diagonal from the right (rd); green to a vertical
direction (v); light blue to a diagonal from the left (ld); and pink to a horizontal force from {the} left (lh).}
\label{fig:IPC}
\end{figure}
IPCs are commercial devices that apply mechanical stress to a section of
an optical fiber to create optical birefringence. The stress is induced by a contact pad that
applies external pressure to the fiber via a screw or actuator. Here we tested a
commercially available device from Thorlabs \cite{Thorlabs_PolarizationController_2014}, originally designed for
single-mode fiber (SMF), applied to a four-core MCF, as illustrated in Fig.~\ref{fig:IPC}. 
The IPC device used here is designed to
apply pressure to the outside of a 900\,$\mu$m diameter tight-buffered fiber. The MCF patchcord
(Fibercore SM-4C1500(8.0/125)/001) used was composed of four cores placed within a 125\,$\mu$m
diameter cladding material, which is covered by a coating material (Dual
Acrylate) reaching 245\,$\mu$m diameter, all within the buffer material \cite{fibercore_datasheet}.

The IPC was placed a few cm from the connector of the patch cord to minimize any twisting of the
fiber, such that the orientation of the cores at the IPC device was very close to that of the FC/PC connector, as shown in
Fig.~\ref{fig:IPC}(b). While the usual use of the device involves applying birefringence and rotating the clamped section
of the fiber around the central axis, here we focus instead on the stress-induced birefringence on
each core of the MCF, so that the only degrees
of freedom are the force applied and the angle of orientation with respect to a certain core
geometry.
\par
IPCs change the polarization state of light through stress-induced birefringence produced by mechanical pressure on a section of the fiber \cite{johnson1979inline,smith1980pressure}.   The stress-optic effect produces a birefringence \cite{Frocht1941Photoelasticity}
\begin{equation}
    \Delta n = C\,(\sigma_1-\sigma_2).
    \label{eq:birefringence}
\end{equation}
Here $C = C_2 - C_1$ is the relative stress-optic coefficient, the difference of
the parallel ($C_1$) and perpendicular ($C_2$) coefficients that set the index change for
light polarized along and across a given stress axis. 
While the birefringence depends only
on $C$, the absolute phase accumulated by each polarization depends on $C_1$ and $C_2$
separately.  

Over a length $L$ of stressed fiber this
gives a retardance (i.e. a differential phase between the slow and fast axes)
$\delta = k L\,\Delta n$,
where $k$ is the wavenumber and $L$ the length of the region under stress. This retardance $\delta$
is precisely the differential phase $\delta=\phi_h-\phi_v$ that appears in the Jones matrix
Eq.~\eqref{eq:Pgen} below. 
The terms $\sigma_1$ and $\sigma_2$ are the principal stresses, corresponding to the eigenvalues of the stress matrix
\begin{equation}
\mathbf{\sigma}=\begin{pmatrix}
        \sigma_{xx} & \sigma_{xy} \\
        \sigma_{yx} & \sigma_{yy} 
    \end{pmatrix}.
    \label{eq:sigma}
\end{equation}
The eigenvectors of matrix \eqref{eq:sigma} determine the direction of the fast and slow optical axes.  Typically, point pressure on a cylinder results in a compressive stress in the direction of the pressure and a tensile stress in the perpendicular direction.  In principle, at each transverse point $(x,y)$ in the fiber,  a corresponding set of principal axes and retardance $\Delta n(x,y)$ is obtained. 
\par
\begin{figure}
{\includegraphics[width=\linewidth]{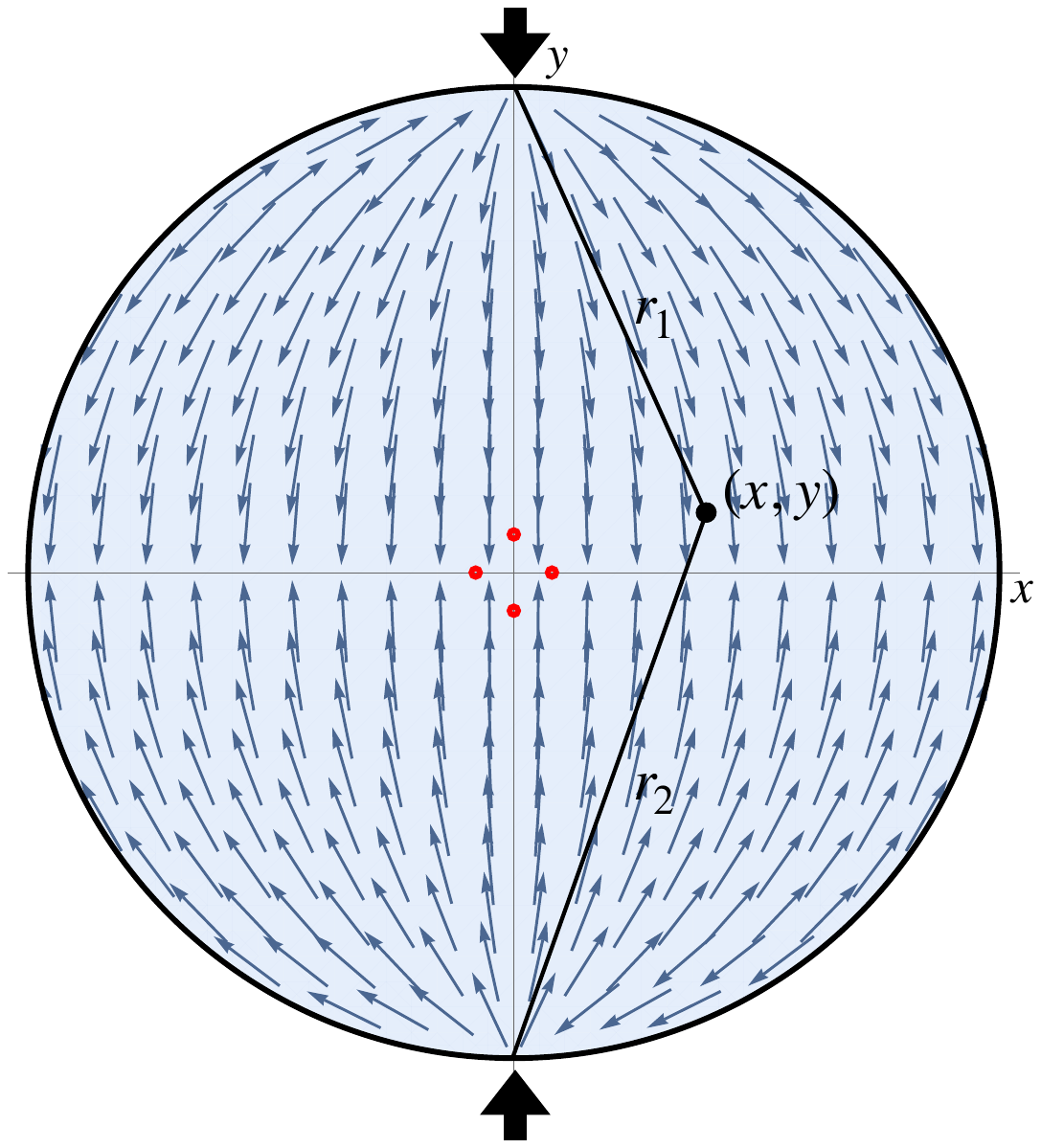}}
\caption{\label{fig:vectorfieldsmall} Principal axes of compressive stress for an optical fiber with
force applied along the vertical ($y$) direction.  The red
circles correspond to positions of the MCF cores. The simplified model assumes a homogeneous fiber material. The arrows are included to aid visualization only.}
\end{figure}

In search of the simplest model for the stress within the MCF, we ignore the different material characteristics of the jacket, buffer and cladding of the fiber, as well as the actual geometry of the contact points of the IPC (a flat contact pad of $\sim 15$mm length on one side, and cylindrical section on the other). Rather, we consider that the IPC applies force at two opposite points of the homogeneous fiber, so that we can model its action by  considering a diametrically compressed disk (see Ref. \cite{gianino81}, or Example 8-10 in Ref. \cite{sadd2005elasticity}), which gives
\begin{equation}
\sigma_{xx} = -\frac{2f}{\pi} \left( \frac{(R - y)x^2}{r_1^4} + \frac{(R + y)x^2}{r_2^4} - \frac{1}{2R} \right)
\label{eq:sigmaxx}
\end{equation}

\begin{equation}
\sigma_{yy} = -\frac{2f}{\pi} \left( \frac{(R - y)^3}{r_1^4} + \frac{(R + y)^3}{r_2^4} - \frac{1}{2R} \right)
\label{eq:sigmayy}
\end{equation}

\begin{equation}
\sigma_{xy} = \frac{2f}{\pi} \left( \frac{(R - y)^2 x}{r_1^4} - \frac{(R + y)^2 x}{r_2^4} \right),
\label{eq:sigmaxy}
\end{equation}
where $f$ is the applied force per unit length, defined as $f=F/L$, with $F$ representing the total applied force and $L$ the contact length. Here by symmetry we have $\sigma_{yx}=\sigma_{xy}$ and the distances from the points of contact to point $(x,y)$
\begin{equation}
r_1 = \sqrt{x^2 + (R - y)^2}, \qquad r_2 = \sqrt{x^2 + (R + y)^2}. 
\end{equation}
Figure \ref{fig:vectorfieldsmall} shows the principal axes of compressive stress where the load is applied in the vertical direction.  These define the fast and slow axes of birefringence at each point in the fiber.  
Also shown in the figure are the principal axes corresponding to compressive stress as a function of the position ($x,y$).  
\par

The resulting effect of the IPC depends greatly upon the geometry of the cores and their radial
position $r=\sqrt{x^2+y^2}$ relative to $R$.  For the IPC considered here, the
dimensions are such that we can consider the small-$r$ limit $r \ll R$, 
where the radial position of the cores is much smaller than the pressure point radius
$R$ ($r \ll R$), we expand Eqs. \eqref{eq:sigmaxx}-\eqref{eq:sigmaxy} to leading order in $x/R$, $y/R$, which gives
\begin{equation}
\sigma_{xx} = \frac{f}{\pi R} \left(1-\frac{4x^2}{R^2} \right),
\label{eq:sigmaxx_s}
\end{equation}
\begin{equation}
\sigma_{yy} = -\frac{f}{\pi R} \left(3+\frac{4y^2}{R^2}-\frac{8x^2}{R^2} \right),
\label{eq:sigmayy_s}
\end{equation}
\begin{equation}
\sigma_{xy} = \frac{f}{\pi R} \left(\frac{8x y}{R^2} \right),
\label{eq:sigmaxy_s}
\end{equation}
where we note that the first correction is second order in $r/R$ (via $x$, $y$). For the jacketed
fiber used in this work, $R=450\,\mu$m (half the 900\,$\mu$m jacket diameter at the IPC contact point) and
$r=25\sqrt{2}\approx 35\,\mu$m (the distance from the fiber axis to each core, given the
50\,$\mu$m square core spacing \cite{fibercore_datasheet}). The second-order terms 
are at most about 2.5\% of the leading terms at the four core positions, and can be safely ignored. Thus, the stress tensor for
all cores reduces to that of a single-mode fiber (SMF) with core at the origin:
\begin{equation}
\sigma_{\mathrm{SMF}}=\left(
\begin{array}{cc}
 \frac{f}{\pi  R} & 0 \\
 0 & -\frac{3 f}{\pi  R} \\
\end{array}
\right),
\label{eq:sigmaSMF}
\end{equation}
where the eigenvectors correspond to the $x$ (perpendicular to direction of force) and $y$
(parallel to direction of force) directions. 

Let us consider first that the IPC is aligned to apply force in the vertical ($y$) direction.
Using Eq.~\eqref{eq:sigmaSMF}, the effect of the IPC on the polarization can be described by the
Jones matrix
\begin{equation}
\oper{P}(0) =
\left (
\begin{matrix}
e^{i \phi_h}  & 0 \\
0 &  e^{i \phi_v}
\end{matrix}
\right ),
\label{eq:P0}
\end{equation}
written in the basis defined by the horizontal ($H$) and vertical ($V$) polarization directions.
The phases $\phi_h$ and $\phi_v$ depend on the force applied by the IPC. We can apply a rotation
matrix to Eq.~\eqref{eq:P0} to write the Jones matrix describing the IPC aligned at angle $\theta$,
giving:
\begin{equation}
\oper{P}(\theta,\delta)=
\left(
\begin{array}{cc}
 e^{i \delta} \cos ^2\theta+\sin ^2\theta & \sin \theta \cos \theta  \left(e^{i\delta}-1\right) \\
 \sin \theta \cos \theta \left(e^{i \delta}-1\right) & e^{i\delta} \sin ^2\theta + \cos ^2 \theta \\
\end{array}
\right),
\label{eq:Pgen}
\end{equation}
where we define the differential phase $\delta=\phi_h-\phi_v$. 
Without loss of generality, we use Dirac notation to describe
the polarization state of light passing through the fiber. Considering an initial state $|\psi_0\rangle$,
the final polarization state is
\begin{equation}\label{rho}
\ket{\psi_f}=\oper{P}(\theta,\delta)|\psi_0\rangle.
\end{equation}
The output polarization state can be measured experimentally via the Stokes parameters, which are
defined in terms of the entries of the density matrix $\rho=\ket{\psi_f}\bra{\psi_f}$ as
\begin{align}
S_1&=\rho_{11}-\rho_{22}\,,\notag\\
S_2&=2\mathrm{Re}(\rho_{12})\,,\label{Sk}\\
S_3&=2\mathrm{Im}(\rho_{21})\,.\notag
\end{align}
The theoretical evolution of the Stokes vector $\vec{S}=(S_1,S_2,S_3)$ for right-circular and left-circular input polarization states are shown in Fig.~\ref{fig:mcfR}.
Expressions
for the Stokes parameters for several common input states and IPC orientations are given in
Appendix~\ref{sec:appendix}.

\subsection{Intra-core and {inter-core} phases}
\label{sec:phases}
The birefringent stress described above governs the polarization transformation within a
single core. In the MCF, the stress experienced by core~$j$ is the cladding stress field evaluated
at the core center,
\begin{equation}
   \sigma^{(j)}_{ab} \;\equiv\; \sigma_{ab}(\mathbf{r}_j), \qquad a,b\in\{x,y\}.
   \label{eq:corestress}
\end{equation}
Let us consider stress applied in the vertical ($y$) direction, and that the geometry is such that two separate cores share common principal stress directions. This holds for the small-$r$ condition, as well as cores aligned along the horizontal or vertical directions in Fig. \ref{fig:vectorfieldsmall}. For light polarized along the horizontal ($H$, $x$) and vertical ($V$, $y$) direction, the
absolute phase accumulated by core $j$ over the stressed length $L$ is set by the stress-optic
indices,
\begin{align}
   \phi_{jh} &= kL\,\big(n_0 - C_1\sigma^{(j)}_{xx} - C_2\sigma^{(j)}_{yy}\big), \notag\\
   \phi_{jv} &= kL\,\big(n_0 - C_1\sigma^{(j)}_{yy} - C_2\sigma^{(j)}_{xx}\big),
   \label{eq:absphase}
\end{align}
where $C_1$ ($C_2$) is the stress-optic coefficient for light polarized parallel (perpendicular) to
a stress axis.  The intra-core retardance is the difference of these,
\begin{equation}
   \delta_j = \phi_{jh}-\phi_{jv} = kL\,(C_2-C_1)\,(\sigma^{(j)}_{xx}-\sigma^{(j)}_{yy}),
   \label{eq:retardance}
\end{equation}
consistent with Eq.~\eqref{eq:birefringence}.
This is the birefringent phase $\delta$ that appears in Eq.~\eqref{eq:Pgen}, and can be accessed
experimentally by polarimetry on each core (Sec.~\ref{sec:exp}).

When the IPC acts on two cores, the relevant {inter-core} observable is the difference of absolute
phases between cores~$i$ and~$j$ for a fixed input polarization, $\varphi_s = \phi_{is}-\phi_{js}$
($s=h,v$). Writing $\Delta\sigma_{ab} = \sigma^{(i)}_{ab}-\sigma^{(j)}_{ab}$ for the core-$i$/core-$j$
stress difference, Eq.~\eqref{eq:absphase} gives
\begin{align}
   \varphi_h &= -kL\,\big(C_1\,\Delta\sigma_{xx}+C_2\,\Delta\sigma_{yy}\big), \notag\\
   \varphi_v &= -kL\,\big(C_2\,\Delta\sigma_{xx}+C_1\,\Delta\sigma_{yy}\big).
   \label{eq:intercore}
\end{align}
These separate naturally into a polarization-independent (common) part and a polarization-dependent
(birefringent) part. Defining 
$\Sigma=\Delta\sigma_{xx}+\Delta\sigma_{yy}$ and $D=\Delta\sigma_{xx}-\Delta\sigma_{yy}$,
\begin{align}
   \varphi_h &= -\frac{kL}{2}\big([C_1+C_2]\Sigma - (C_2-C_1)D\big), \notag\\
   \varphi_v &= -\frac{kL}{2}\big([C_1+C_2]\Sigma + (C_2-C_1)D\big).
   \label{eq:intercore2}
\end{align}
Equation~\eqref{eq:intercore2} is the central structural result of the model. The first term is
the {inter-core} phase common for both input polarizations. The second term is the birefringent part, governed
by the difference $C_2-C_1$ and by the stress-anisotropy difference $D$. The half-sum
$\tfrac{1}{2}(\varphi_h+\varphi_v) = -\tfrac{kL}{2}(C_1+C_2)\Sigma$ isolates the first, and the
difference isolates the second,
\begin{equation}
   \varphi_h-\varphi_v = kL\,(C_2-C_1)\,D \;=\; \delta_1-\delta_4.
   \label{eq:consistency}
\end{equation}
The last equality holds for
any mechanism, photoelastic or otherwise, provided only that $H$ and $V$ are principal axes at both
cores.  The two observables are complementary: the intra-core retardance $\delta_j$
[Eq.~\eqref{eq:retardance}] is measured by polarimetry, while the {inter-core} phase $\varphi_s$
[Eq.~\eqref{eq:intercore}] is accessible by two-core interferometry
(Sec.~\ref{sec:interferometry}). Eq. \eqref{eq:consistency} therefore provides an alternative route to probing the relative retardance imparted to different cores. 

A remark on the geometry is in order for what follows. For diametric compression along $y$, the exact solution
Eqs.~\eqref{eq:sigmaxx}--\eqref{eq:sigmaxy} gives $\sigma_{xx}(0,y)=f/\pi R$ independent of $y$,
$\sigma_{yy}(0,y)$ even in $y$, and $\sigma_{xy}(0,y)=0$. The model predicts identical
stress tensors, so $\Sigma = D = 0$ and both {inter-core} phases vanish. Thus, {measuring} any difference in intra-core birefringence or {inter-core} phase in cores 1 and 4 is an indication of the departure of the real device from the simplistic model.   

\section{Experiment}
\label{sec:exp}
\begin{figure}
{\includegraphics[width=\linewidth]{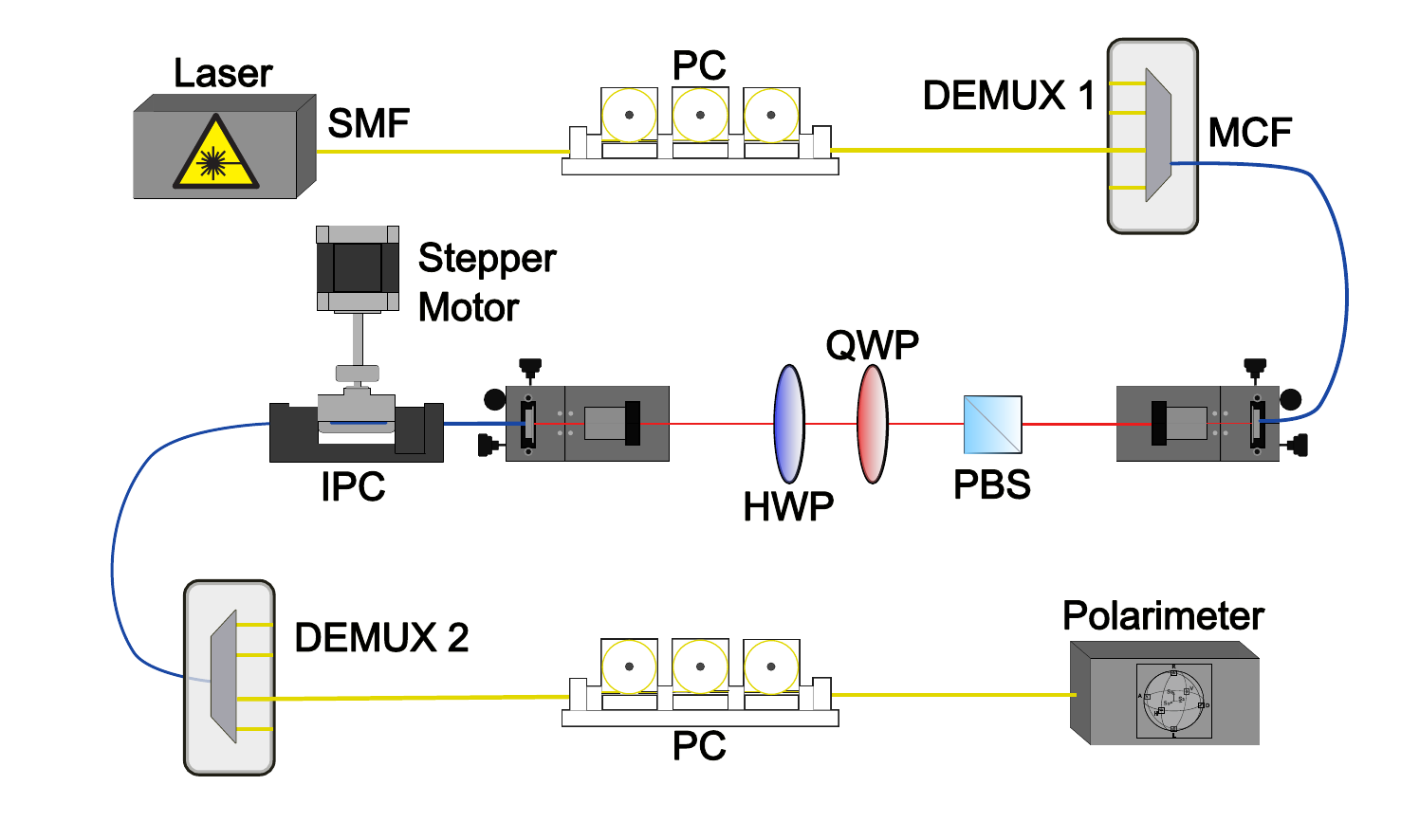}}
\caption{Set-up for the characterization of the polarization in multi-core fiber. PC: polarization
controller, SMF: single-mode fiber, MCF: multi-core fiber, DEMUX: fiber (de)multiplexer for single-core
SMF to MCF, PBS: polarizing beam splitter, QWP: quarter-wave plate, HWP: half-wave plate.}
\label{fig:setup}
\end{figure}
\subsection{Single-core Polarimetry}
The experimental setup for the characterization of the IPC on the four-core MCF is shown in
Fig.~\ref{fig:setup}. A 1550~nm continuous-wave laser sends light through a single-mode fiber
(SMF), which is connected to a fiber (de)multiplexer (DEMUX) to couple the signal into a multi-core
fiber (MCF). A polarization controller (PC) is used on the SMF to optimize the input polarization
(see below) before coupling the light into free space using an objective lens.

The free-space section is used to set the input polarization and to guarantee that it is the same
for all cores of the fiber. A polarizing beam splitter (PBS) followed by a quarter-wave plate
(QWP) and a half-wave plate (HWP) are used to prepare the same polarization state $|\psi_{in}
\rangle$ at the entrance to all MCF cores at the MCF connector. We can thus ignore all polarization
transformations occurring before the polarizer, though the fiber PC was used to optimize
transmission. After the free-space polarization optics, the light passes through another objective
lens which forms an image of the exit face of the previous MCF on a second MCF patchcord (5~m
length), as has been done in previous work \cite{taddei21}. The IPC is placed immediately after
the FC/PC connector, to ensure that the cores are arranged close to the ``+'' geometry shown in
Fig.~\ref{fig:IPC}, and to minimize any fiber-induced polarization changes due to propagation in
the fiber. The IPC implements the transformation $\oper{P}(\theta,\delta)$ on core $j$, where
$\theta$ is the direction along which the mechanical stress is applied. After the IPC, the light
travels through the MCF patchcord, through a DEMUX into a SMF and then to a polarimeter. A
3-paddle polarization controller $\oper{M}$ is used on the MCF patchcord to correct polarization
rotations due to propagation within the MCF. The output polarization state going into the
polarimeter can be described by
\begin{equation}
|\psi \rangle = \oper{M}_j \oper{U}_j \oper{P}(\theta,\delta) |\psi_{in} \rangle,
\end{equation}
where $\oper{U}_j$ is the polarization transformation due to propagation through MCF core $j$, as
well as the DEMUX and SMF (losses are ignored). Here $\oper{M}_j$ is the transformation due to the
PC acting on core $j$. Ideally, when testing core $j$, we can adjust
$\oper{M}_j=\oper{U}_j^{-1}$, so that
\begin{equation}
|\psi_j \rangle = \oper{P}(\theta,\delta) |\psi_{in} \rangle.
\end{equation}
This was done in an initial polarization calibration stage by preparing several non-orthogonal
input states {(e.g. $H$, $V$, $R$, $L$)} and adjusting the PC so that the corresponding states were observed at the polarimeter.

\label{sec:polarimetry_dk}
Polarimetry measurements were performed for all cores with the IPC aligned in five different
directions: vertical, two horizontal orientations, and two diagonal orientations, as shown in
Fig.~\ref{fig:IPC}(b). A NEMA~17 stepper motor at quarter-microstep resolution (800~pulses per revolution) was coupled to the actuator screw and used to control the force applied by the IPC.
The motor was configured to complete one revolution every 10~s. To ensure
consistent mechanical pressure during measurements, the number of screw turns was recorded starting from a pre-defined zero position.
Polarization changes appear around turn~12, and measurements were conducted from turn~10 to
turn~15.5. No further pressure beyond turn~15.5 was applied to the fiber to avoid the risk of
damage to the patchcord.

\begin{figure*}
{\includegraphics[width=\linewidth]{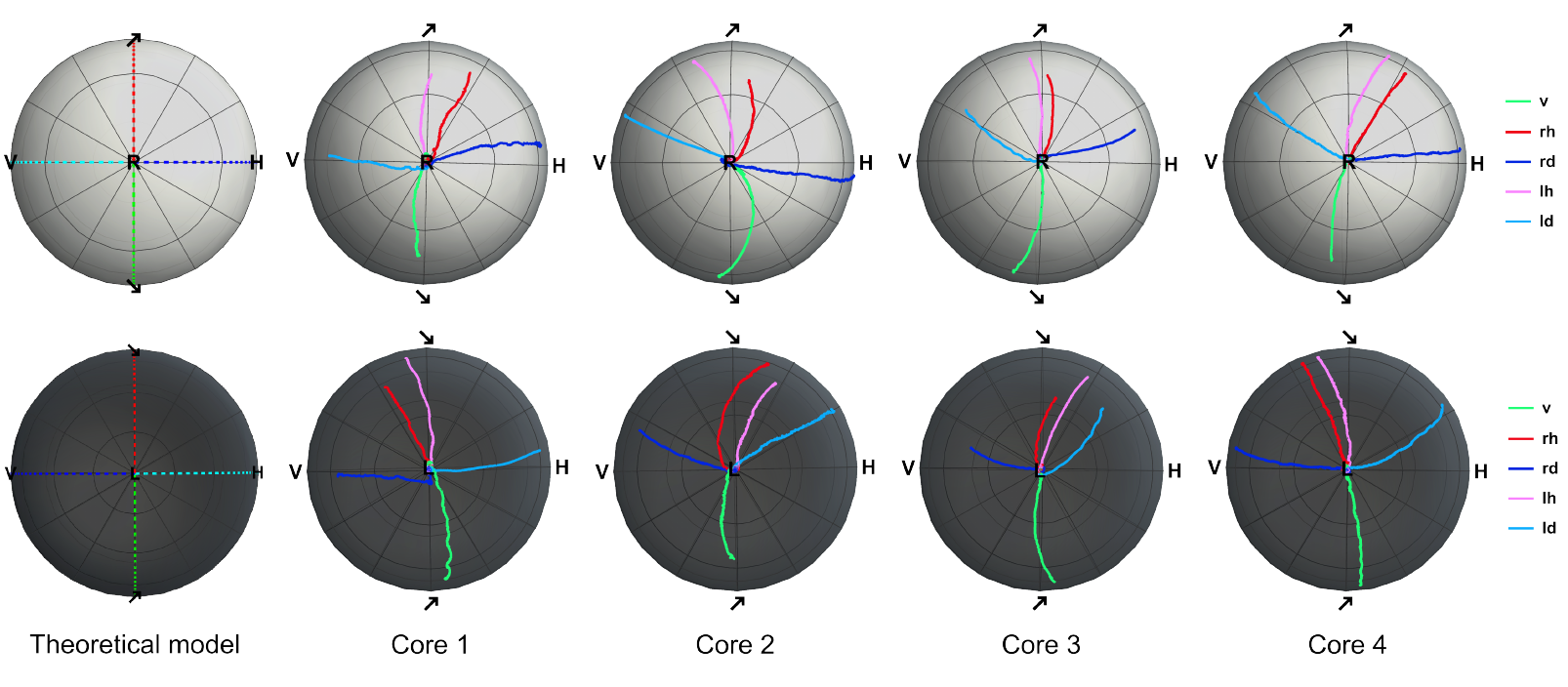}}
\caption{Bloch sphere trajectories of the polarization state for all four MCF cores under IPC
action, for an initial right-circular (R) input polarization (top row) and left-circular (L) polarization (bottom row). The leftmost sphere shows the
theoretical prediction based on the small-$r$ model (Sec.~\ref{sec:ipc}), which is the same for
all cores. {Colors indicate the IPC orientation: green -- vertical (v); blue -- diagonal right
(rd); light blue -- diagonal left (ld); red -- horizontal right (rh); pink -- horizontal left
(lh).} Deviations from the theoretical trajectory are attributed primarily to residual
fiber-induced polarization rotations in the $\sim$11~m of MCF, DEMUX, and SMF after the IPC (see
text).}
\label{fig:mcfR}
\end{figure*}

The polarimetry results are shown in Fig.~\ref{fig:mcfR}  for initial
right-circular and left-circular polarization states. The Bloch spheres for each
figure show the trajectory of the polarization state as the fiber is acted on by the IPC. We also
show the trajectory corresponding to the theoretical model based on the small-$r$ approximation
given in Sec.~\ref{sec:ipc}, which applies to all cores. The agreement is qualitative
rather than quantitative. All trajectories run from the circular pole toward the linear states, as
the model requires, but they depart visibly from the predicted great circle and from one another,
following curves of smaller radius that do not close on the equator at the predicted point. This is due primarily to the fiber-induced polarization rotations that occur in the MCF
patch cord, DEMUX and SMF ($\sim$11~m length in total) after the IPC. Though these were minimized
using the PC, even small deviations from $\oper{M}_j = \oper{U}_j^{-1}$ lead to considerable
changes in the trajectory of the polarization state. A secondary contribution might arise from movement
of the glass fiber inside the jacket material during the IPC actuation, though we consider this
effect to be minimal in comparison to the residual fiber rotations.

The second physically relevant question concerns the magnitude of the retardance in each core.  Two additional 
limitations affect our ability to make a quantitative comparison via single-core polarimetry. First we observed 
run-to-run variation between separately acquired data sets, even for similar operating conditions.  This could be due to hysteresis and mechanical memory of the jacket and/or buffer material. Second, imprecision in determining the zero position of force applied by the tension screw adds additional uncertainty.  Thus, for the polarization measurements we can make rough estimate of the magnitude of the retardance, and resort to interferometric measurements for the {inter-core} comparison. 

A linear retarder acts on
the Poincar\'e sphere as a rotation through the retardance $\delta$, and the circular and linear
states are separated by $90^\circ$, so the fraction of the polar distance traversed gives a rough estimate of the
retardance directly off the sphere. Every core in Fig.~\ref{fig:mcfR} covers at least half of that
distance, placing $\delta_j \gtrsim 45^\circ$ for all four. This is a conservative reading: the
measured trajectories depart from the great circle expected for an input exactly at the pole, and
for a tilted rotation axis a given polar displacement corresponds to a larger retardance than the
great-circle reading implies. By comparison, applying the same measure to a pre-onset window of
equal length, in which the actuator contacts the jacket but does not yet load the fiber, returns
between $0.6^\circ$ and $5^\circ$, and no response is seen below turn~12 in any run. Each core
therefore acquires on the order of one radian of birefringent phase, and does so only when the actuator
is loading the fiber.

{Independently, the right- and left-circular input states constitute two
measurements of the same intra-core retardance, and they agree to a median of $13^\circ$, which is due most likely to the fact that measurements are performed independently. We
therefore expect a resolution of no better than $\sim 15^\circ$ for retardance
\emph{differences} between cores, which prevents us from predicting any meaningful core-to-core retardance differences  using the polarimetry data.   Moreover, this prohibits testing any of the reflection symmetry predicted by the simplified model for our specific core geometry, which is most likely violated by the actual asymmetric device. It also prohibits verification of predicted relation of Eq.~\eqref{eq:consistency}, which
requires core-to-core resolution well below the value quoted above. We expect that more quantitative statements could be  made using a bare fiber and simultaneous polarimetry on cores.}

\begin{figure}
\includegraphics[width=\linewidth]{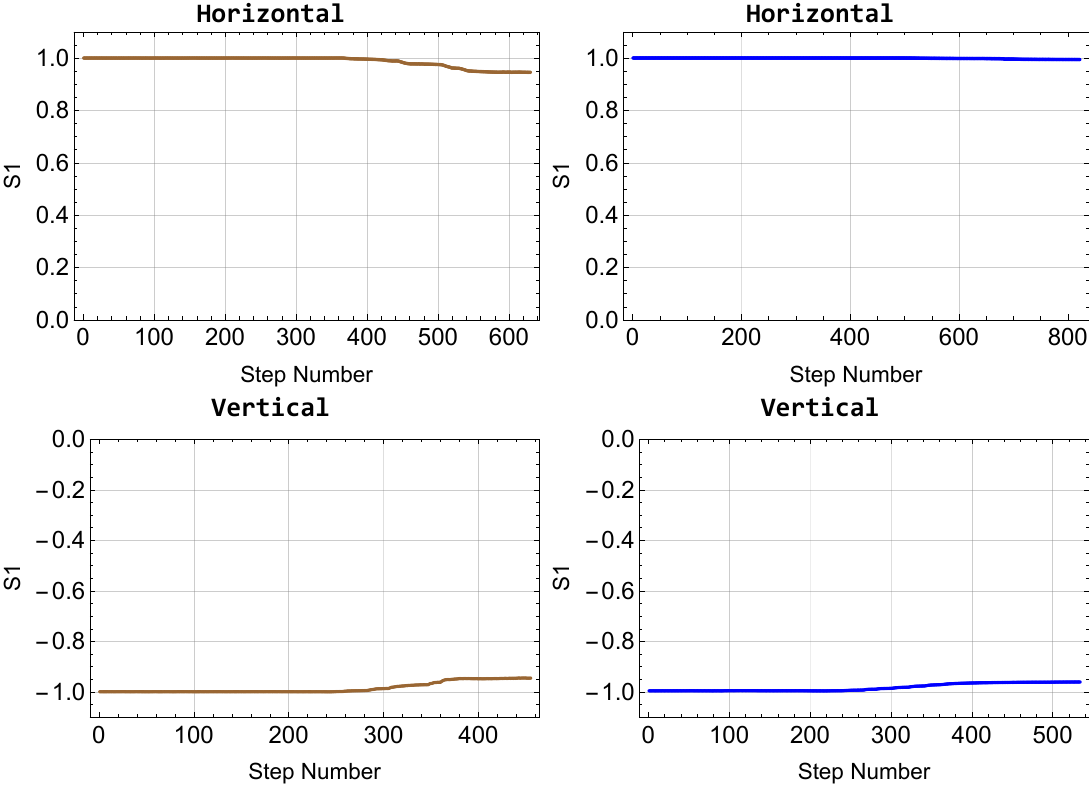}
\caption{Stokes parameter $S_1$ as a function of screw step for core 1 (Brown) and core 4 (Blue), with horizontal and vertical input polarization under
vertical IPC pressure. In both cases $S_1$ remains nearly constant, confirming that the IPC acts
as a pure phase retarder when the input polarization is aligned with the principal stress axis.
Discontinuities are due to manual (non-motorized) application of pressure during these particular
measurements.}
\label{fig:stokes_HV}
\end{figure}

For the interferometric measurement reported in section \ref{sec:interferometry}, it is essential that the IPC act only on the {inter-core}
phase, without rotating the polarization within each core, which would itself modulate the observed
fringe. When the input polarization is along a
principal stress axis ($H$ or $V$ for $\theta = 0$) the Jones matrix Eq.~\eqref{eq:Pgen} is diagonal in the $H/V$ basis,
so each core acquires an absolute phase while its polarization state is unchanged. {This holds for any value of the retardance, so the operating condition is set by the input polarization alone, and not by the screw position or by the details of the stress model.}
Figure~\ref{fig:stokes_HV} confirms this, with $S_1 \approx +1$ ($H$ input) and $S_1 \approx -1$
($V$ input) held constant across the actuator range for both cores~1 and~4. Cores~1 and~4 therefore
acquire different absolute phases at fixed polarization, and the resulting {inter-core} phase is read
out directly in the output intensities. {The residual polarization rotations downstream of the IPC, which are responsible for the departure of the trajectories in Fig.~\ref{fig:mcfR} from the model, do not affect this condition. Because the interferometer contains no polarizing element and these rotations are static, they can at most change the static fringe visibility, which is included in the fringe range $2I_0$ measured with the motor off (Sec.~\ref{sec:interferometry}).}

\subsection{Two-core Interferometry}
\label{sec:interferometry}

\begin{figure}
{\includegraphics[width=\linewidth]{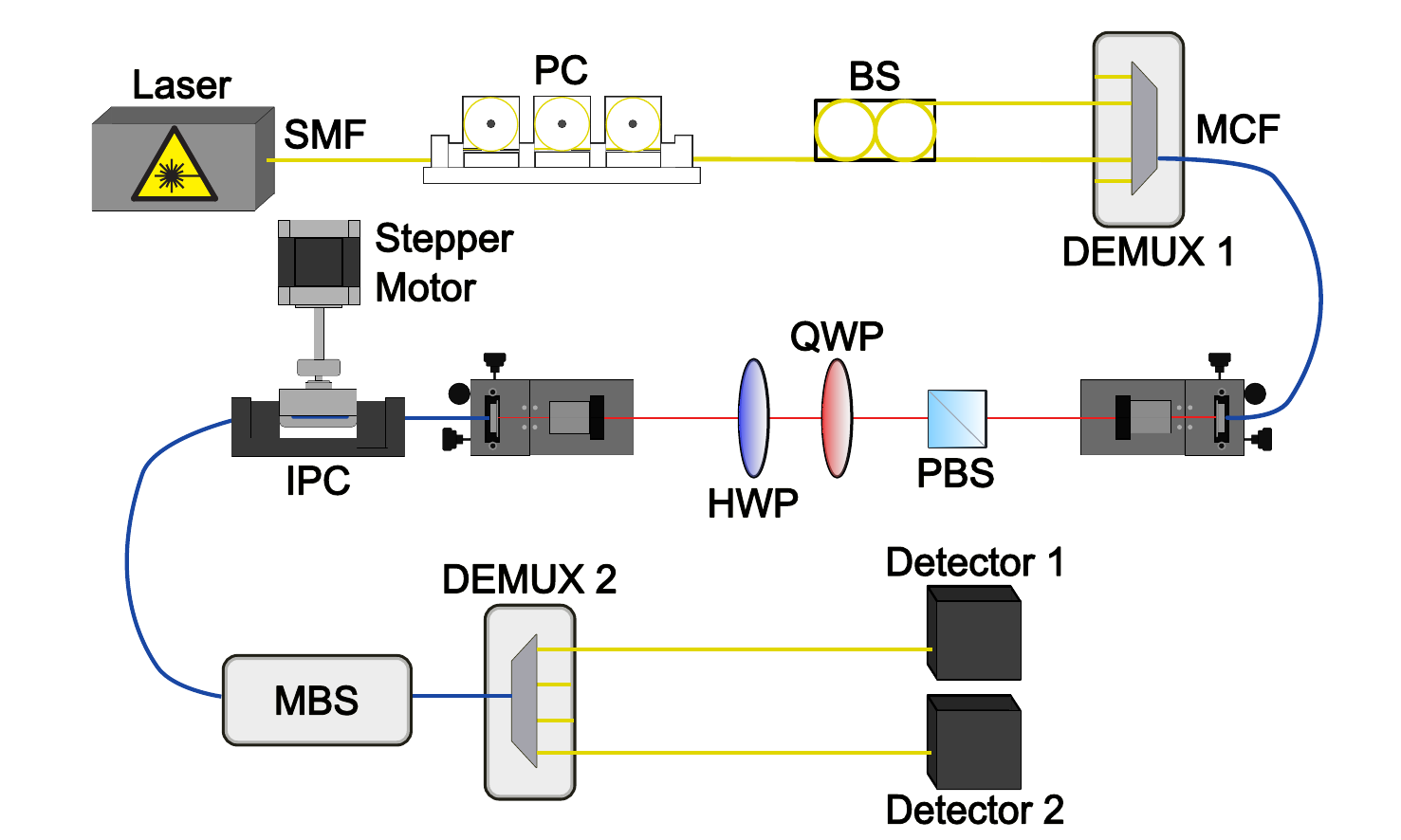}}
\caption{{Experimental setup for the two-core interferometric phase measurements. PC:
polarization controller, SMF: single-mode fiber, MCF: multi-core fiber, DEMUX: fiber (de)multiplexer, PBS: polarizing beam splitter, QWP: quarter-wave plate, HWP: half-wave plate, BS: 2$\times$2 fiber beam splitter, MBS: multi-core fiber beam splitter.}}
\label{fig:setup2}
\end{figure}

To probe the {inter-core} phase
directly,  a two-core interferometer was constructed, as shown in
Fig.~\ref{fig:setup2}. The setup is similar to Fig.~\ref{fig:setup}, but with a 2$\times$2 fiber
beam splitter (BS) used to divide the incoming laser light into cores~1 and~4, and a multi-core
fiber beam splitter (MBS) \cite{carine20} used to recombine the light, which was then sent into
the DEMUX and SMFs. Two single-photon detectors (ID Qube series, efficiency 10\%, dead time 5\,$\mu$s) were used to record photocounts at the two output ports of the MBS while force from
the IPC was applied. The use of single-photon detectors required attenuation of the input
laser, and before each measurement run the total count rates were set to approximately 30\,000 counts per second \footnote{Though the measurement is one of classical first-order (intensity) interference, the single-photon detectors were the instruments readily available, and operating at the single-photon level also confirms that the inline phase control remains effective in the low-flux
regime relevant to the quantum-photonic applications partially motivating this work.}.  The {inter-core} phases $\varphi_h$ and $\varphi_v$ for H and V input polarizations can then be extracted from the fringe data.

%{A limitation of this setup is that the 2$\times$2 fiber BS is placed before the MCF connector, meaning the two optical paths to cores~1 and~4 travel separately through the BS and connecting fiber before recombination. This introduces extra phase noise from environmental fluctuations in the separate paths, on top of the IPC-induced phase. To distinguish the IPC-driven phase modulation from this background noise, the data are analyzed in the frequency domain, as described below.}
\subsubsection{Time-domain oscillations}
\label{sec:timedomain}
\begin{figure}
\includegraphics[width=\linewidth]{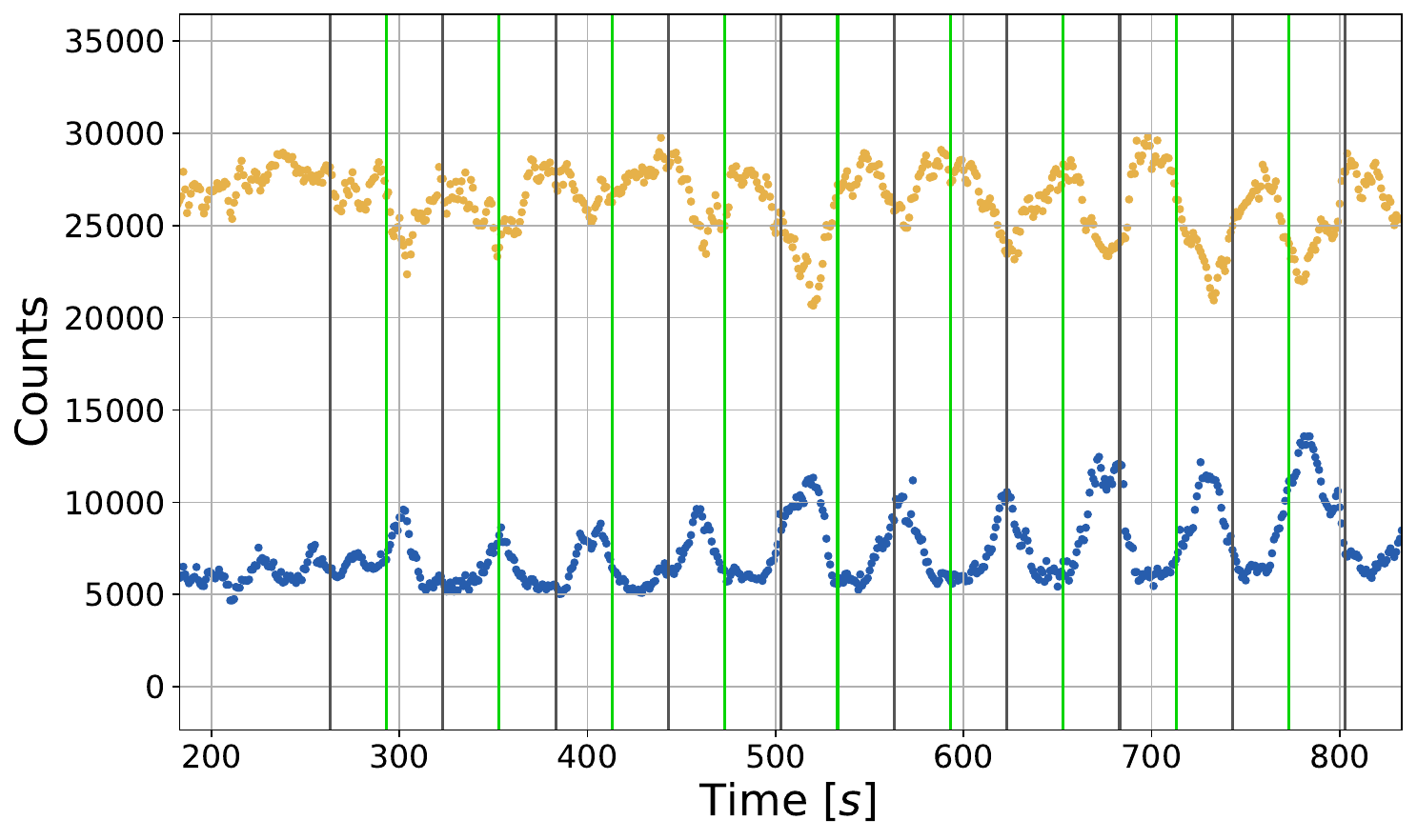}
\caption{Photon count rates at detectors D1 (blue) and D2 (orange) as a function of time,
with vertical input polarization in cores~1 and~4 and the IPC applying vertical pressure
(motorized actuator cycling through three revolutions/30s and back). The
anti-correlated oscillations between the two output ports of the MBS are the interference fringes
produced by the IPC-induced {inter-core} phase modulation. Deviations are due to phase drift from ambient effects.}

\label{fig:timeseries_V}
\end{figure}

\begin{figure}
\includegraphics[width=\linewidth]{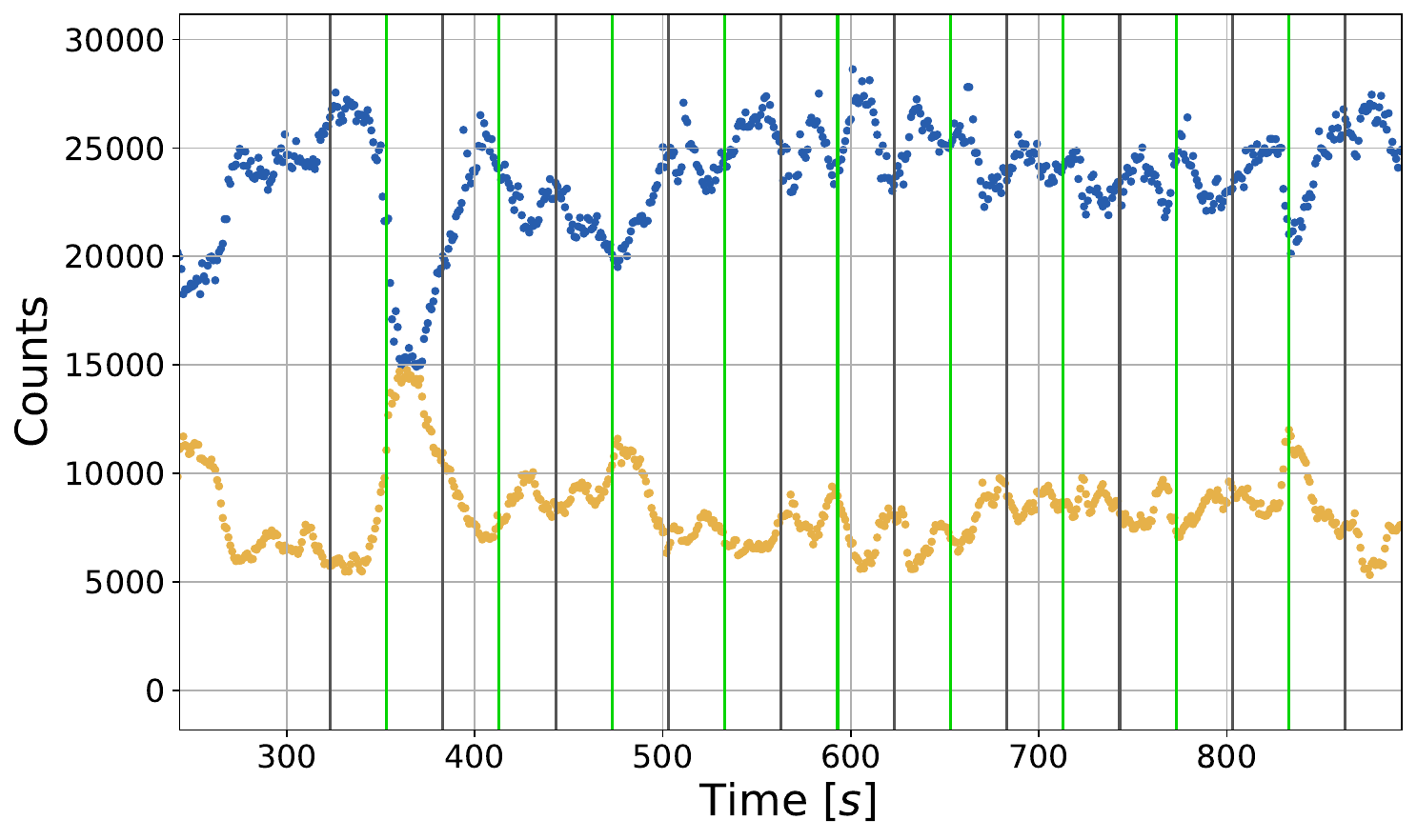}
\caption{Same as Fig.~\ref{fig:timeseries_V} but with horizontal input polarization.
The oscillation amplitude is observably smaller than for V polarization (compare
Fig.~\ref{fig:timeseries_V}), which also makes the phase drift more noticeable.}
\label{fig:timeseries_H}
\end{figure}

Figures~\ref{fig:timeseries_H} and~\ref{fig:timeseries_V} show representative time traces of the photon counts at detectors D1 and D2 for H and V input polarizations, respectively. In both cases,  oscillations are visible and the two detectors are anti-correlated, as expected. These anti-correlated oscillations are interference fringes between the light from cores~1 and~4 at the MBS, driven by the IPC-induced inter-core phase.
The oscillation amplitude is consistently larger for V than for H input. The IPC screw is ramped in
and out over each $60\,$s cycle and the fringe follows at $f_\mathrm{mot}$ over several successive
cycles, so the response is parabolic in shape resulting from the repeated ramp-in/ramp-out cycle. 

{Inspection of the time traces shows that
successive fringe extrema grow slowly over a run, with later cycles reaching larger amplitudes than
earlier ones, which is consistent with viscoelastic settling of the polymer jacket and coating under
repeated load. Mechanical memory of this kind is expected for a polymer-buffered fiber and we have
not characterized it here. It is, however, largely irrelevant if the IPC is used for stabilization of the sub-hertz {inter-core} phase drift of an all-MCF interferometer, where the actuator would be
driven by a feedback loop tracking an error signal. {With regards to reproducibility, the observed modulation frequency followed the actuator frequency at all three motor speeds tested (see below). In the polarimetry runs, no response was seen below turn~12. In contrast, the screw position at which the response begins is not reproducible, because the force zero cannot be set precisely, and the fringe extrema grow slowly over a run. We therefore do not claim open-loop repeatability of the phase. The intended closed-loop use requires only that the inter-core phase increase monotonically with the applied force, which is consistent with the parabolic fringe traces of Fig.~\ref{fig:timeseries_V}.}}

{Visible in the time traces is also the irregular background modulation in the count rates due to the phase instability introduced by the 2$\times$2
SMF BS that splits the laser beam before coupling into cores~1 and~4, which occurs on timescales comparable to the IPC cycling period. To separate the IPC-driven signal from this background, we performed a spectral analysis using the fast Fourier transform.}

\subsubsection{Frequency analysis and phase extraction}

\begin{figure}
% Mark the expected motor frequency f_mot = 0.016 Hz and the measured peak.
\includegraphics[width=\linewidth]{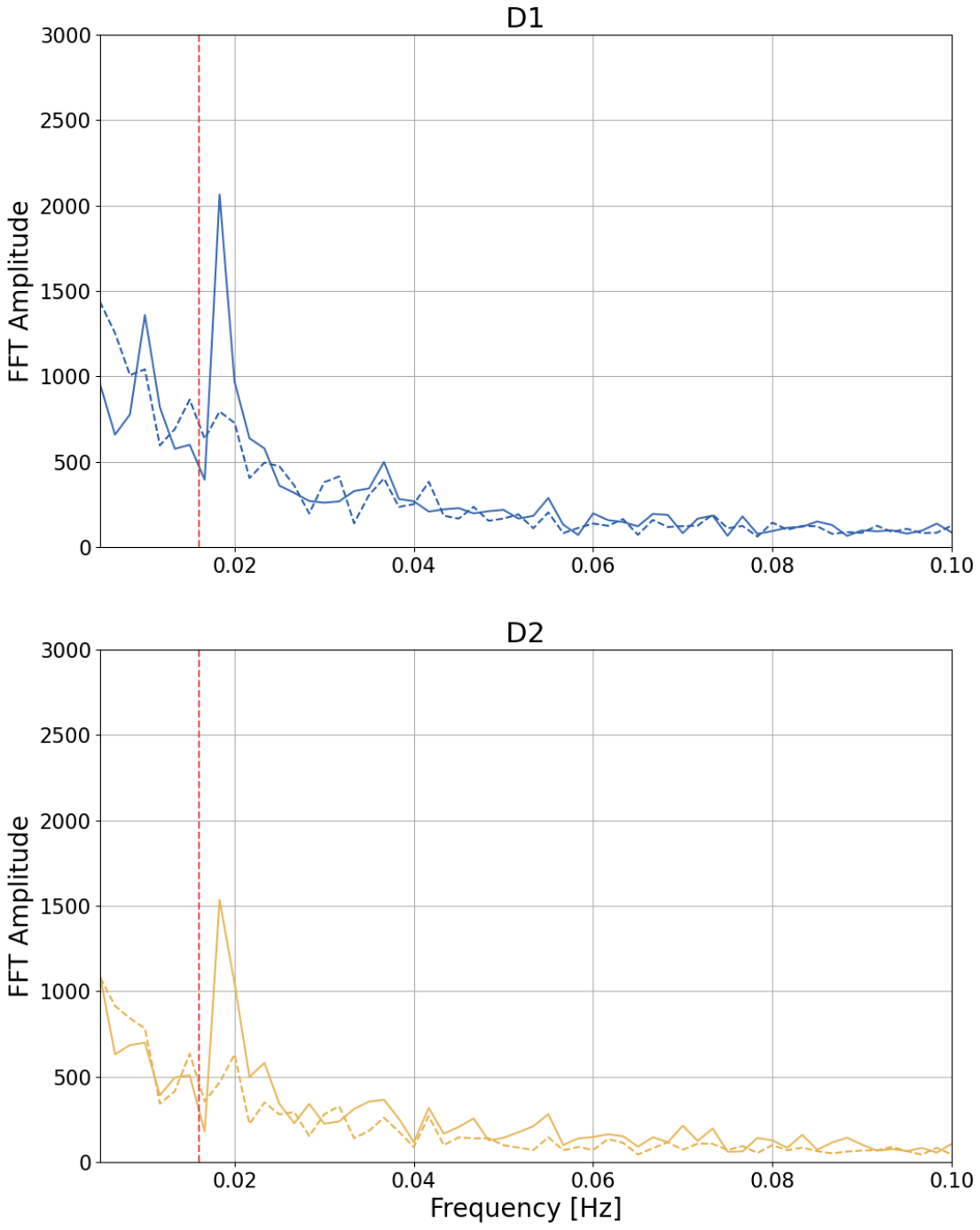}
\caption{Power spectrum of the photon-count oscillations at detectors D1 (top figure) and D2 (bottom figure), averaged
over all runs, for $V$ input (solid lines) and $H$ input (dashed lines). The expected motor frequency
$f_\mathrm{mot} = 1/60\,\mathrm{Hz} \approx 0.016$\,Hz (one full screw load/unload cycle in
$60\,$s) is indicated by the vertical red dashed line. The V-polarization spectrum shows a clear peak
near $f_\mathrm{mot}$, while the H-polarization peak is barely above the natural phase fluctuations of the interferometer. }
\label{fig:fft_HV}
\end{figure}

{Figure~\ref{fig:fft_HV} shows the power spectra obtained via Fast Fourier Transform (FFT) of the time traces averaged over all
repeated runs for each polarization. A clear peak appears near the expected motor frequency
$f_\mathrm{mot} = 1/60\,\mathrm{Hz} \approx 0.016$\,Hz for V input, and a smaller peak for H
input. The slight shift from $f_\mathrm{mot}$ to the slightly higher $\sim 0.018$ is most likely due to errors from our homemade stepper motor controller.  A small residual drift of the operating
phase point $\varphi_0$ over the measurement most likely contributes to the peak width through a weak amplitude
modulation of the fringe, though $\varphi_0$ was approximately constant over the few cycles used
here. The peak amplitude provides a cleaner, more noise-robust estimate of the IPC-induced fringe amplitude $A$, separated from the low-frequency phase
noise previously mentioned.}
The spectra sit on a sloped background that is largest at low frequency and falls toward higher
frequency, the signature of the slow environmental phase drift in the separate SMFs before the first DMUX. The
V-input peak stands well above this background and its height is consistent with the oscillation
amplitude seen directly in the time trace, whereas the H-input peak is comparable to the local
background level, so the H amplitude is best regarded as an upper bound (see below).

To confirm that the modulation is driven by the IPC rather than by ambient drift, we repeated the measurement at different motor speeds. Cycling the screw at $5$, $10$, and $15\,$s/revolution moves the drive frequency in proportion, and the fringe peak follows: the $15\,$s/rev data peak near $0.013\,\mathrm{Hz}$ (expected $1/90\,\mathrm{Hz}\approx 0.011\,\mathrm{Hz}$) and the $5\,$s/rev data near $0.038\,\mathrm{Hz}$ (expected $1/30\,\mathrm{Hz}\approx 0.033\,\mathrm{Hz}$). The peaks therefore lie 8 to 17\% above the corresponding $f_\mathrm{mot}$, including the $10\,$s/rev data, a shift that we attribute to the stepper-motor controller. {The spectral analysis of the phase modulation is consistent across different motor speeds, and the fiber showed no physical damage after multiple measurements.} The $10\,$s/rev rate used for the data reported here gave the best tradeoff between the fiber's limited mechanical bandwidth at faster rates and increasing phase-drift contamination at slower rates.

To extract the {inter-core} phase difference from the fringe amplitude, we use the two-port
interference relation for a single detector,
\begin{equation}
    I = \bar{I} + I_0 \cos\varphi,
    \label{eq:interference}
\end{equation}
where $\bar{I}$ is the mean count rate, $I_0$ the fringe amplitude (so the full fringe peak-to-peak
is $2I_0$), and $\varphi$ the {inter-core} phase difference. Before each run the static phase was set to an easily reproducible operating point by adjusting the input BS fibers to minimize
the count rate at one detector, corresponding to $\varphi_0\approx\pi$. As the IPC screw drives the {inter-core} phase from 0 to $\delta \varphi$, the count rate
responds quadratically, as can be seen in Fig. \ref{fig:timeseries_V}. The
peak-to-peak modulation of the count rate is
\begin{equation}
   G = I_0\,\left(1-\cos\delta\varphi\right),
   \label{eq:ppswing}
\end{equation}
where $2I_0$ is the full fringe peak-to-peak count range, obtained with the
motor off while the environmental phase drifts freely over several oscillation periods, and reported in Table~\ref{tab:phases}.

Equation~\eqref{eq:ppswing} relates the {inter-core} phase to $G$, the peak-to-peak modulation of
the count rate driven by the IPC. However, the spectral analysis gives
$A_1$, the amplitude of the Fourier component at $f_\mathrm{mot}$, and the two differ by a factor
that depends on the shape of the driven waveform. The IPC pad advances and retracts linearly in
time, so the {inter-core} phase follows a triangular ramp between $0$ and $\delta\varphi$. At the operating point $\varphi_0 = \pi$, the count
rate responds quadratically, giving a parabolic wave form. Computing its Fourier fundamental gives
\begin{equation}
   A_1 = I_0\,\frac{2\,\delta\varphi\,\sin\delta\varphi}{\pi^2-\delta\varphi^2},
   \label{eq:fundamental}
\end{equation}
which we solve numerically for $\delta\varphi$. In the small-phase limit Eq.~\eqref{eq:fundamental} reduces to
$A_1 = (4/\pi^2)\,G \simeq 0.405\,G$, and it departs from that limit by less than $1.5\%$ over the
range of phases measured here, so the ratio $A_1/G$ is essentially constant.

Equation~\eqref{eq:ppswing} assumes that the background phase
$\varphi_0$ is a multiple of $\pi$. A conservative bound that is \emph{independent} of the
operating point follows by noting that, for a fixed count difference $G$, the smallest possible phase
modulation occurs at $\varphi_0=\pi/2$, giving
\begin{equation}
   \delta\varphi \;\geq\; \delta\varphi_{\mathrm{min}} = 2\arcsin\!\left(\frac{G}{2I_{0}}\right),
   \label{eq:phase_lb}
\end{equation}
with $G = A_1\pi^2/4$ from the shape factor above.
Equations~\eqref{eq:fundamental} and~\eqref{eq:phase_lb} are applied independently to each
detector and each input polarization.

The resulting {inter-core} phase modulation depths are summarized in
Table~\ref{tab:phases}. For $V$ input the extraction gives $\delta\varphi = 0.95$ and $0.90$~rad at
the two detectors ($54.7^\circ$ and $51.3^\circ$), with an operating-point-independent lower bound
above $0.38$~rad. The values of $\delta\varphi$ assume that the static phase was set to $\varphi_0\approx\pi$ before each run. For $H$ input the values are $0.58$ and $0.59$~rad. The $H$ spectral peak,
however, sits at a height comparable to the local drift background (Fig.~\ref{fig:fft_HV}), so
$A_1$ for $H$ is biased upward by the background and the $H$ entries must be read as upper bounds
rather than as measurements with the quoted precision.

\begin{table}
\caption{\label{tab:phases} {inter-core} phase modulation depths extracted from the mean of
three independent experimental runs, for the two detectors (D1, D2) and both input polarizations
(V, H). $A_1$ is the amplitude of the spectral peak at $f_\mathrm{mot}$ and $2I_0$ is the full
fringe interference range. The value $\delta\varphi$ follows from Eq.~\eqref{eq:fundamental}
and the operating-point-independent lower bound $\delta\varphi_{\min}$ from
Eq.~\eqref{eq:phase_lb}. Uncertainties are statistical,
propagated from $A_1$ and $2I_0$. The $H$ entries are background-limited and thus are taken as upper bounds.}
\begin{ruledtabular}
\begin{tabular}{lcc}
 & \textbf{D1} & \textbf{D2} \\ \hline
\multicolumn{3}{l}{Vertical (V)} \\ \hline
$A_1$ (counts)                    & 2063.5 $\pm$ 66.6    & 1534.5 $\pm$ 78.8     \\
$2I_0$ (counts)  & 23733.3 $\pm$ 185.2  & 19939.6 $\pm$ 176.3   \\
$\delta\varphi$ (deg)           & 54.7 $\pm$ 1.0 & 51.3 $\pm$ 1.4  \\
$\delta\varphi$ (rad)           & 0.95 $\pm$ 0.02 & 0.90 $\pm$ 0.02  \\
$\delta\varphi_{\min}$ (deg)    & $24.4 \pm 0.4$           & $21.6 \pm 0.6$            \\ \hline
\multicolumn{3}{l}{Horizontal (H), upper bounds} \\ \hline
$A_1$ (counts)                    & $\leq$ 794.0 $\pm$ 88.5     & $\leq$ 629.7 $\pm$ 58.9      \\
$2I_0$ (counts)  & 23924.0 $\pm$ 187.5  & 18079.8 $\pm$ 170.8   \\
$\delta\varphi$ (deg)           & $\leq$ 33.2 $\pm$ 1.9 & $\leq$ 34.0 $\pm$ 1.6  \\
$\delta\varphi$ (rad)           & $\leq$ 0.58 & $\leq$ 0.59  \\
$\delta\varphi_{\min}$ (deg)    & $\leq 9.3\pm0.5$            & $\leq 9.7\pm0.5$             \\
\end{tabular}
\end{ruledtabular}
\end{table}

We note that polarization modulation from the IPC cannot account for the observed modulation {for $V$ input}. The interferometer
contains no polarizing element, so a mismatch between the polarization states of the two cores can only reduce the fringe visibility,
$\mathcal{V} = \sqrt{\tfrac{1}{2}(1+\vec{S}^{(1)}\!\cdot\vec{S}^{(4)})} = |\cos(\Theta/2)|$, with
$\Theta$ the angle between the two Stokes vectors. At an extremum a change in visibility displaces
the count rate toward the mean by $(1-\mathcal{V})I_0$. The polarimetry of
Fig.~\ref{fig:stokes_HV} gives $|S_1| \geq 0.95$ for each core throughout the actuation.  Assuming the extreme case of opposite rotation (which is contrary to the common evolution seen in
Fig.~\ref{fig:mcfR}), gives $\Theta \leq 36^\circ$ and $\mathcal{V}_{\min} = 0.95$, and a bound of
$0.05\,I_0$ on any purely polarization-driven effect. {For $V$ input the driven modulation depth is $G/I_0 = 0.43$, which exceeds this worst-case bound by a factor of $8.6$. For $H$ input, $G/I_0\leq 0.16$ is itself only an upper bound, dominated by the background, so the comparison is not meaningful and we use of the $H$ entries of Table~\ref{tab:phases} only as upper bounds on the inter-core phase.}

As a final, preliminary test we applied two IPCs in series, modulated in unison. The fringe
amplitude grew: from the $V$-polarization time trace we obtain $A_1 \approx 3000$ and a full
peak-to-peak $2I_0 \approx 16000$ counts, giving $\delta\varphi \approx 1.5$~rad ($\approx
84^\circ$), roughly $1.5$ times the single-IPC value. This
measurement was considerably noisier than the single-IPC data, plausibly from slight asynchrony between the two independently
driven actuators which were place side by side. We therefore present it only as preliminary evidence that the phase accumulates
across devices. In principle, there is no reason why
concatenation of IPC devices should not increase the imprinted phase, so 
 that phases of $\pi$ or more may be within reach.

\subsection{Physical interpretation and limits of the model}
\label{sec:compare}

The polarimetric and interferometric measurements access complementary information. The dominant,
polarization-independent part of the {inter-core} phase is governed by $\Sigma$, the difference in
mean transverse stress between the cores. A term of this form shifts $n_h$ and $n_v$ together at a
given core, so it produces no retardance for any input polarization and is inaccessible to
polarimetry.  

It is worth being explicit about what the appearance of an {inter-core} phase implies. For
diametric compression along $y$, the solution of Sec.~\ref{sec:ipc} gives
$\sigma_{xx}(0,y)$ independent of $y$, $\sigma_{yy}(0,y)$ even in $y$, and $\sigma_{xy}(0,y)=0$, so
cores~1 and~4 see identical stress tensors and the {inter-core} phase vanishes. This is exact, to all
orders in $r/R$, and follows from the mirror symmetry of the idealized loading rather than from any
approximation. The measurement is therefore a null test of that idealization, and the 
$\approx 1$~rad phase difference observed for V polarization is a direct measure of the departure of the real device, a flat pad bearing on one
side of a jacketed fiber seated in a cylindrical bore, from the symmetric configuration assumed in
the model. Characterizing that geometry is a separate problem from the one addressed here.

{Beyond the model, the measured ordering of the phases may give some insight into the physical origin of the observed phase difference. Squaring and subtracting Eqs.~\eqref{eq:intercore}, the cross terms cancel and give the exact relation
\begin{equation}
   \varphi_v^2-\varphi_h^2 = (kL)^2\,(C_2^2-C_1^2)\,(\Delta\sigma_{xx}^2-\Delta\sigma_{yy}^2).
   \label{eq:ordering}
\end{equation}
Since $|C_2|>|C_1|$ [$C_1=0.65\times10^{-12}$~Pa$^{-1}$ and $C_2=4.2\times10^{-12}$~Pa$^{-1}$ \cite{ocampo25}], the observation $|\varphi_v|>|\varphi_h|$ implies $|\Delta\sigma_{xx}|>|\Delta\sigma_{yy}|$, so that the inter-core stress difference is greater in the direction transverse to the loading axis than along it. This conclusion rests on three assumptions: the inter-core phase is entirely photoelastic (so that polarization-independent contributions, such as a path-length difference due to bending, are absent), that both cores experience the same stressed length $L$, and that $H$ and $V$ are principal stress axes at both cores.  The noise limit of the $H$ channel does not weaken the ordering, provided that the $H$ and $V$ runs were taken with the same static phase $\varphi_0\approx\pi$, and because the background can only bias the $H$ amplitude upward.  Given these assumptions and the limited precision of the $H$ data, this interpretation should be regarded as indicative only, and a more precise measurement of $\varphi_h$ would be required to confirm it.}

{Finally, one consequence matters directly for use of the device. Writing the measured phases as a common part and a polarization-dependent part, and assuming that $\varphi_h$ and $\varphi_v$ have the same sign, the latter amounts to $(\varphi_v-\varphi_h)/(\varphi_v+\varphi_h)\approx 0.2$ of the former ($0.24$ and $0.21$ for detectors D1 and D2, respectively). Because $\varphi_h$ is only an upper bound, this ratio is a lower bound on the polarization-dependent fraction. For $V$ input, at most about 80\% of the imparted phase is therefore independent of the input polarization. The IPC is thus not fully polarization independent, and the true polarization dependence may be larger than estimated here. In an application where the input polarization is neither known nor fixed, this residual dependence sets the accuracy with which the inter-core phase can be controlled. Operating with the input aligned to a principal axis, as here, removes the ambiguity entirely.}

\section{Conclusions}
\label{sec:conclusions}

We have shown that a commercial inline polarization controller functions as an inline
{inter-core} phase shifter for {multi-core} fiber. Two-core interferometry gives a reversible {inter-core}
phase of $0.95 \pm 0.02$~rad for vertical input polarization, with an operating-point-independent
lower bound of $0.38$~rad, and at most $0.59$~rad for horizontal input. The screw is ramped in and
out over each cycle and the fringe follows the drive frequency across several successive cycles and
across three actuation rates.

We emphasize that here we demonstrate the repurposing of a device for phase control rather than
characterization of stress-induced retardance in the MCF. An IPC is normally adjusted by
rotating the clamped section about the fiber axis while squeezing, so that the birefringence axis
is steered and the fiber twisted.  Here the rotational degree of freedom instead fixes the load axis
at a chosen angle relative to the core lattice, a setting that has no meaning for single-mode fiber, and the
force alone is varied. The result is an off-the-shelf, low-loss, inline component that
delivers close to a radian of {inter-core} phase without fanout to SMF or free-space optics.  We observe mechanical memory effects, even under sub-hertz modulation. 
However, as the intended application is not high-speed modulation but
stabilization of the slow, sub-hertz {inter-core} phase drift of an all-MCF interferometer, any slowly varying  mechanical response can be
absorbed by a tracking loop{, provided that the inter-core phase increases monotonically with the applied force.} 

A simple diametrically compressed disk model is used to predict that for an input polarization along a principal stress axis determined by the Jones
matrix, each core acquires an absolute phase at fixed polarization and the device
acts as a pure {inter-core} phase shifter. 
The model ignores key characteristics of the IPC and fiber, and is used as a guide only to identify the operating
conditions. Indeed, at leading order it predicts no {inter-core} phase at
all between the pair of cores measured. Thus the observed signal is a direct indication of the deviation of the real contact
geometry from the simplified model. We expect that each device and mounting will behave somewhat
differently, similar to the paddle polarization controllers typically used in fiber optics. 

There are several avenues open for future work. The phase range can plausibly be extended: the screw travel limit was
not reached in any measurement, and preliminary data with two devices in series reach
$\approx 1.5$~rad. Faster actuation would require replacing the manual IPC with a piezoelectric or
electro-optic polarization squeezer while retaining the geometry.  Also interesting is to apply the same load to
a bare, unjacketed MCF, where the stress components then vary appreciably across the core
positions, and the
Jones matrix becomes core dependent. This is the regime already exploited for multiaxis load and
bend sensing with Bragg gratings in bare MCF~\cite{silva-lopez2004,flockhart03}. Used as an
actuator rather than a sensor, and with the load applied at an angle neither along nor
perpendicular to the core axes, it would couple the spatial and polarization degrees of freedom of
the guided light within the fiber itself, opening a route to combined path-polarization operations
in an all-fiber geometry.

\appendix
\section{Stokes parameters for special cases}
\label{sec:appendix}
The Stokes parameters~\eqref{Sk} follow directly from Eqs.~\eqref{eq:Pgen} and~\eqref{rho}
for arbitrary $\theta$ and $\delta$. Table~\ref{tab:Stokes} collects the results for the six
cardinal input states at the IPC orientations used in this work. For $\theta = 0$ the first two rows
confirm that $H$ and $V$ are eigenstates of $\oper{P}(0,\delta)$ and remain unchanged under vertical
pressure for any $\delta$, which is the pure-phase-shifter condition exploited in
Sec.~\ref{sec:interferometry}. The remaining rows show that the diagonal and circular inputs evolve
with a sensitivity that grows with $\delta$, which is why circular input is used for the polarimetry
of Fig.~\ref{fig:mcfR}.

\begin{table*}[t]
\centering{
\renewcommand{\arraystretch}{1.3}
\begin{tabular}{|c|c|c|c|c|}
\hline
States & $\theta = 0$ & $\theta = \pi/2$ & $\theta = \pi/4$ & $\theta = -\pi/4$ \\ \hline
(H) & $(1,0,0)$ & $(1,0,0)$ & $(\cos\delta, 0, -\sin\delta)$ & $(\cos\delta, 0, \sin\delta)$ \\ \hline
(V) & $(-1,0,0)$ & $(-1,0,0)$ & $(-\cos\delta, 0, \sin\delta)$ & $(-\cos\delta, 0, -\sin\delta)$ \\ \hline
(D) & $(0,\cos\delta,-\sin\delta)$ & $(0,\cos\delta,\sin\delta)$ & $(0, 1, 0)$ & $(0, 1, 0)$ \\ \hline
(A) & $(0,-\cos\delta,\sin\delta)$ & $(0,-\cos\delta,-\sin\delta)$ & $(0, -1, 0)$ & $(0, -1, 0)$ \\ \hline
(R) & $(0,\sin\delta,\cos\delta)$ & $(0,-\sin\delta,\cos\delta)$ & $(\sin\delta, 0, \cos\delta)$ & $(-\sin\delta, 0, \cos\delta)$ \\ \hline
(L) & $(0,-\sin\delta,-\cos\delta)$ & $(0,\sin\delta,-\cos\delta)$ & $(-\sin\delta, 0, -\cos\delta)$ & $(\sin\delta, 0, -\cos\delta)$ \\ \hline
\end{tabular}}
\caption{Theoretical Stokes vectors $(S_1, S_2, S_3)$ as a function of the IPC orientation angle
$\theta$, for horizontal (H), vertical (V), diagonal (D), anti-diagonal (A), right-circular (R),
and left-circular (L) input polarization.}
\label{tab:Stokes}
\end{table*}

\begin{acknowledgments}
This research was funded by ANID FONDECYT-Regular Grants No. (1230796,  1231940, 1230897, 1240746, 1260111), ANID  Millennium Science Initiative Program  ICN17$_-$012,  ANID Anillo Project ATE250003, and ANID AC3E CIA250006.
\end{acknowledgments}


\begin{thebibliography}{38}%
\makeatletter
\providecommand \@ifxundefined [1]{%
 \@ifx{#1\undefined}
}%
\providecommand \@ifnum [1]{%
 \ifnum #1\expandafter \@firstoftwo
 \else \expandafter \@secondoftwo
 \fi
}%
\providecommand \@ifx [1]{%
 \ifx #1\expandafter \@firstoftwo
 \else \expandafter \@secondoftwo
 \fi
}%
\providecommand \natexlab [1]{#1}%
\providecommand \enquote  [1]{``#1''}%
\providecommand \bibnamefont  [1]{#1}%
\providecommand \bibfnamefont [1]{#1}%
\providecommand \citenamefont [1]{#1}%
\providecommand \href@noop [0]{\@secondoftwo}%
\providecommand \href [0]{\begingroup \@sanitize@url \@href}%
\providecommand \@href[1]{\@@startlink{#1}\@@href}%
\providecommand \@@href[1]{\endgroup#1\@@endlink}%
\providecommand \@sanitize@url [0]{\catcode `\\12\catcode `\$12\catcode
  `\&12\catcode `\#12\catcode `\^12\catcode `\_12\catcode `\%12\relax}%
\providecommand \@@startlink[1]{}%
\providecommand \@@endlink[0]{}%
\providecommand \url  [0]{\begingroup\@sanitize@url \@url }%
\providecommand \@url [1]{\endgroup\@href {#1}{\urlprefix }}%
\providecommand \urlprefix  [0]{URL }%
\providecommand \Eprint [0]{\href }%
\providecommand \doibase [0]{https://doi.org/}%
\providecommand \selectlanguage [0]{\@gobble}%
\providecommand \bibinfo  [0]{\@secondoftwo}%
\providecommand \bibfield  [0]{\@secondoftwo}%
\providecommand \translation [1]{[#1]}%
\providecommand \BibitemOpen [0]{}%
\providecommand \bibitemStop [0]{}%
\providecommand \bibitemNoStop [0]{.\EOS\space}%
\providecommand \EOS [0]{\spacefactor3000\relax}%
\providecommand \BibitemShut  [1]{\csname bibitem#1\endcsname}%
\let\auto@bib@innerbib\@empty
%</preamble>
\bibitem [{\citenamefont {Xavier}\ and\ \citenamefont {Lima}(2020)}]{xavier20}%
  \BibitemOpen
  \bibfield  {author} {\bibinfo {author} {\bibfnamefont {G.~B.}\ \bibnamefont
  {Xavier}}\ and\ \bibinfo {author} {\bibfnamefont {G.}~\bibnamefont {Lima}},\
  }\href {https://doi.org/10.1038/s42005-019-0269-7} {\bibfield  {journal}
  {\bibinfo  {journal} {Communications Physics}\ }\textbf {\bibinfo {volume}
  {3}},\ \bibinfo {pages} {9} (\bibinfo {year} {2020})}\BibitemShut {NoStop}%
\bibitem [{\citenamefont {Richardson}(2010)}]{richardson10}%
  \BibitemOpen
  \bibfield  {author} {\bibinfo {author} {\bibfnamefont {D.~J.}\ \bibnamefont
  {Richardson}},\ }\href {https://doi.org/10.1126/science.1191708} {\bibfield
  {journal} {\bibinfo  {journal} {Science}\ }\textbf {\bibinfo {volume}
  {330}},\ \bibinfo {pages} {327} (\bibinfo {year} {2010})},\ \Eprint
  {https://arxiv.org/abs/https://science.sciencemag.org/content/330/6002/327.full.pdf}
  {https://science.sciencemag.org/content/330/6002/327.full.pdf} \BibitemShut
  {NoStop}%
\bibitem [{\citenamefont {Richardson}\ \emph {et~al.}(2013)\citenamefont
  {Richardson}, \citenamefont {Fini},\ and\ \citenamefont
  {Nelson}}]{richardson13}%
  \BibitemOpen
  \bibfield  {author} {\bibinfo {author} {\bibfnamefont {D.~J.}\ \bibnamefont
  {Richardson}}, \bibinfo {author} {\bibfnamefont {J.~M.}\ \bibnamefont
  {Fini}},\ and\ \bibinfo {author} {\bibfnamefont {L.~E.}\ \bibnamefont
  {Nelson}},\ }\href {https://doi.org/10.1038/nphoton.2013.94} {\bibfield
  {journal} {\bibinfo  {journal} {Nature Photonics}\ }\textbf {\bibinfo
  {volume} {7}},\ \bibinfo {pages} {354} (\bibinfo {year} {2013})}\BibitemShut
  {NoStop}%
\bibitem [{\citenamefont {Puttnam}\ \emph {et~al.}(2021)\citenamefont
  {Puttnam}, \citenamefont {Rademacher},\ and\ \citenamefont
  {Lu\'{i}s}}]{Puttnam:21}%
  \BibitemOpen
  \bibfield  {author} {\bibinfo {author} {\bibfnamefont {B.~J.}\ \bibnamefont
  {Puttnam}}, \bibinfo {author} {\bibfnamefont {G.}~\bibnamefont
  {Rademacher}},\ and\ \bibinfo {author} {\bibfnamefont {R.~S.}\ \bibnamefont
  {Lu\'{i}s}},\ }\href {https://doi.org/10.1364/OPTICA.427631} {\bibfield
  {journal} {\bibinfo  {journal} {Optica}\ }\textbf {\bibinfo {volume} {8}},\
  \bibinfo {pages} {1186} (\bibinfo {year} {2021})}\BibitemShut {NoStop}%
\bibitem [{\citenamefont {Melo}\ \emph {et~al.}(2025)\citenamefont {Melo},
  \citenamefont {Reyes~F}, \citenamefont {Arroyo}, \citenamefont {G{\'o}mez},
  \citenamefont {Walborn}, \citenamefont {Lima}, \citenamefont {Figueroa},
  \citenamefont {Cari{\~n}e},\ and\ \citenamefont {Saavedra}}]{melo2025}%
  \BibitemOpen
  \bibfield  {author} {\bibinfo {author} {\bibfnamefont {C.}~\bibnamefont
  {Melo}}, \bibinfo {author} {\bibfnamefont {M.}~\bibnamefont {Reyes~F}},
  \bibinfo {author} {\bibfnamefont {D.}~\bibnamefont {Arroyo}}, \bibinfo
  {author} {\bibfnamefont {E.~S.}\ \bibnamefont {G{\'o}mez}}, \bibinfo {author}
  {\bibfnamefont {S.~P.}\ \bibnamefont {Walborn}}, \bibinfo {author}
  {\bibfnamefont {G.}~\bibnamefont {Lima}}, \bibinfo {author} {\bibfnamefont
  {M.}~\bibnamefont {Figueroa}}, \bibinfo {author} {\bibfnamefont
  {J.}~\bibnamefont {Cari{\~n}e}},\ and\ \bibinfo {author} {\bibfnamefont
  {G.}~\bibnamefont {Saavedra}},\ }\href
  {https://doi.org/10.1038/s44172-025-00412-7} {\bibfield  {journal} {\bibinfo
  {journal} {Communications Engineering}\ }\textbf {\bibinfo {volume} {4}},\
  \bibinfo {pages} {77} (\bibinfo {year} {2025})}\BibitemShut {NoStop}%
\bibitem [{\citenamefont {Deakin}\ \emph {et~al.}(2019)\citenamefont {Deakin},
  \citenamefont {Enrico}, \citenamefont {Parsons},\ and\ \citenamefont
  {Zervas}}]{BeamSteering}%
  \BibitemOpen
  \bibfield  {author} {\bibinfo {author} {\bibfnamefont {C.}~\bibnamefont
  {Deakin}}, \bibinfo {author} {\bibfnamefont {M.}~\bibnamefont {Enrico}},
  \bibinfo {author} {\bibfnamefont {N.}~\bibnamefont {Parsons}},\ and\ \bibinfo
  {author} {\bibfnamefont {G.}~\bibnamefont {Zervas}},\ }\href
  {https://doi.org/10.1109/JLT.2019.2896318} {\bibfield  {journal} {\bibinfo
  {journal} {Journal of Lightwave Technology}\ }\textbf {\bibinfo {volume}
  {37}},\ \bibinfo {pages} {1954} (\bibinfo {year} {2019})}\BibitemShut
  {NoStop}%
\bibitem [{\citenamefont {Winzer}\ and\ \citenamefont
  {Neilson}(2017)}]{Winzer:17}%
  \BibitemOpen
  \bibfield  {author} {\bibinfo {author} {\bibfnamefont {P.~J.}\ \bibnamefont
  {Winzer}}\ and\ \bibinfo {author} {\bibfnamefont {D.~T.}\ \bibnamefont
  {Neilson}},\ }\href
  {https://opg.optica.org/jlt/abstract.cfm?URI=jlt-35-5-1099} {\bibfield
  {journal} {\bibinfo  {journal} {J. Lightwave Technol.}\ }\textbf {\bibinfo
  {volume} {35}},\ \bibinfo {pages} {1099} (\bibinfo {year}
  {2017})}\BibitemShut {NoStop}%
\bibitem [{\citenamefont {Saitoh}\ and\ \citenamefont
  {Matsuo}(2016)}]{crosstalkdef}%
  \BibitemOpen
  \bibfield  {author} {\bibinfo {author} {\bibfnamefont {K.}~\bibnamefont
  {Saitoh}}\ and\ \bibinfo {author} {\bibfnamefont {S.}~\bibnamefont
  {Matsuo}},\ }\href {https://doi.org/10.1109/JLT.2015.2466444} {\bibfield
  {journal} {\bibinfo  {journal} {Journal of Lightwave Technology}\ }\textbf
  {\bibinfo {volume} {34}},\ \bibinfo {pages} {55} (\bibinfo {year}
  {2016})}\BibitemShut {NoStop}%
\bibitem [{\citenamefont {Yao}\ \emph {et~al.}(2023)\citenamefont {Yao},
  \citenamefont {Zhao},\ and\ \citenamefont {Tang}}]{yao23}%
  \BibitemOpen
  \bibfield  {author} {\bibinfo {author} {\bibfnamefont {Y.}~\bibnamefont
  {Yao}}, \bibinfo {author} {\bibfnamefont {Z.}~\bibnamefont {Zhao}},\ and\
  \bibinfo {author} {\bibfnamefont {M.}~\bibnamefont {Tang}},\ }\href
  {https://doi.org/10.3390/s23073436} {\bibfield  {journal} {\bibinfo
  {journal} {Sensors}\ }\textbf {\bibinfo {volume} {23}},\ \bibinfo {pages}
  {3436} (\bibinfo {year} {2023})}\BibitemShut {NoStop}%
\bibitem [{\citenamefont {Zhao}\ \emph {et~al.}(2020)\citenamefont {Zhao},
  \citenamefont {Tang},\ and\ \citenamefont {Lu}}]{Zhao2020}%
  \BibitemOpen
  \bibfield  {author} {\bibinfo {author} {\bibfnamefont {Z.}~\bibnamefont
  {Zhao}}, \bibinfo {author} {\bibfnamefont {M.}~\bibnamefont {Tang}},\ and\
  \bibinfo {author} {\bibfnamefont {C.}~\bibnamefont {Lu}},\ }\href
  {https://doi.org/10.29026/oea.2020.190024} {\bibfield  {journal} {\bibinfo
  {journal} {Opto-Electronic Advances}\ }\textbf {\bibinfo {volume} {3}},\
  \bibinfo {pages} {190024} (\bibinfo {year} {2020})}\BibitemShut {NoStop}%
\bibitem [{\citenamefont {Song}\ \emph {et~al.}(2023)\citenamefont {Song},
  \citenamefont {Sun}, \citenamefont {Jiang}, \citenamefont {Chen},
  \citenamefont {Zhu}, \citenamefont {Ren},\ and\ \citenamefont
  {Wang}}]{Song2023}%
  \BibitemOpen
  \bibfield  {author} {\bibinfo {author} {\bibfnamefont {J.}~\bibnamefont
  {Song}}, \bibinfo {author} {\bibfnamefont {S.}~\bibnamefont {Sun}}, \bibinfo
  {author} {\bibfnamefont {C.}~\bibnamefont {Jiang}}, \bibinfo {author}
  {\bibfnamefont {H.}~\bibnamefont {Chen}}, \bibinfo {author} {\bibfnamefont
  {X.}~\bibnamefont {Zhu}}, \bibinfo {author} {\bibfnamefont {J.}~\bibnamefont
  {Ren}},\ and\ \bibinfo {author} {\bibfnamefont {S.}~\bibnamefont {Wang}},\
  }\href {https://doi.org/10.1016/j.yofte.2022.103296} {\bibfield  {journal}
  {\bibinfo  {journal} {Optical Fiber Technology}\ }\textbf {\bibinfo {volume}
  {71}},\ \bibinfo {pages} {103296} (\bibinfo {year} {2023})}\BibitemShut
  {NoStop}%
\bibitem [{\citenamefont {Suo}\ \emph {et~al.}(2022)\citenamefont {Suo},
  \citenamefont {Peng}, \citenamefont {Wang}, \citenamefont {Zhao},\ and\
  \citenamefont {Gao}}]{Suo2022}%
  \BibitemOpen
  \bibfield  {author} {\bibinfo {author} {\bibfnamefont {L.}~\bibnamefont
  {Suo}}, \bibinfo {author} {\bibfnamefont {Y.-P.}\ \bibnamefont {Peng}},
  \bibinfo {author} {\bibfnamefont {M.}~\bibnamefont {Wang}}, \bibinfo {author}
  {\bibfnamefont {W.-Z.}\ \bibnamefont {Zhao}},\ and\ \bibinfo {author}
  {\bibfnamefont {L.-C.}\ \bibnamefont {Gao}},\ }\href
  {https://doi.org/10.3390/s22155630} {\bibfield  {journal} {\bibinfo
  {journal} {Sensors (Basel)}\ }\textbf {\bibinfo {volume} {22}},\ \bibinfo
  {pages} {5630} (\bibinfo {year} {2022})}\BibitemShut {NoStop}%
\bibitem [{\citenamefont {Cuando-Espitia}\ \emph {et~al.}(2021)\citenamefont
  {Cuando-Espitia}, \citenamefont {Fuentes-Fuentes}, \citenamefont
  {Velázquez-Benítez}, \citenamefont {Amezcua}, \citenamefont
  {Hernández-Cordero},\ and\ \citenamefont
  {May-Arrioja}}]{Cuando-Espitia2021}%
  \BibitemOpen
  \bibfield  {author} {\bibinfo {author} {\bibfnamefont {N.}~\bibnamefont
  {Cuando-Espitia}}, \bibinfo {author} {\bibfnamefont {M.~A.}\ \bibnamefont
  {Fuentes-Fuentes}}, \bibinfo {author} {\bibfnamefont {A.}~\bibnamefont
  {Velázquez-Benítez}}, \bibinfo {author} {\bibfnamefont {R.}~\bibnamefont
  {Amezcua}}, \bibinfo {author} {\bibfnamefont {J.}~\bibnamefont
  {Hernández-Cordero}},\ and\ \bibinfo {author} {\bibfnamefont
  {D.}~\bibnamefont {May-Arrioja}},\ }\href
  {https://doi.org/10.1038/s41598-021-97646-0} {\bibfield  {journal} {\bibinfo
  {journal} {Scientific Reports}\ }\textbf {\bibinfo {volume} {11}},\ \bibinfo
  {pages} {18345} (\bibinfo {year} {2021})}\BibitemShut {NoStop}%
\bibitem [{\citenamefont {Cari{\~n}e}\ \emph {et~al.}(2021)\citenamefont
  {Cari{\~n}e}, \citenamefont {Asan-Srain}, \citenamefont {Lima},\ and\
  \citenamefont {Walborn}}]{carine21}%
  \BibitemOpen
  \bibfield  {author} {\bibinfo {author} {\bibfnamefont {J.}~\bibnamefont
  {Cari{\~n}e}}, \bibinfo {author} {\bibfnamefont {M.~N.}\ \bibnamefont
  {Asan-Srain}}, \bibinfo {author} {\bibfnamefont {G.}~\bibnamefont {Lima}},\
  and\ \bibinfo {author} {\bibfnamefont {S.~P.}\ \bibnamefont {Walborn}},\
  }\href {https://doi.org/10.1038/s41534-021-00502-2} {\bibfield  {journal}
  {\bibinfo  {journal} {npj Quantum Information}\ }\textbf {\bibinfo {volume}
  {7}},\ \bibinfo {pages} {172} (\bibinfo {year} {2021})}\BibitemShut {NoStop}%
\bibitem [{\citenamefont {Taddei}\ \emph {et~al.}(2021)\citenamefont {Taddei},
  \citenamefont {Cari\~ne}, \citenamefont {Mart\'{\i}nez}, \citenamefont
  {Garc\'{\i}a}, \citenamefont {Guerrero}, \citenamefont {Abbott},
  \citenamefont {Ara\'ujo}, \citenamefont {Branciard}, \citenamefont {G\'omez},
  \citenamefont {Walborn}, \citenamefont {Aolita},\ and\ \citenamefont
  {Lima}}]{taddei21}%
  \BibitemOpen
  \bibfield  {author} {\bibinfo {author} {\bibfnamefont {M.~M.}\ \bibnamefont
  {Taddei}}, \bibinfo {author} {\bibfnamefont {J.}~\bibnamefont {Cari\~ne}},
  \bibinfo {author} {\bibfnamefont {D.}~\bibnamefont {Mart\'{\i}nez}}, \bibinfo
  {author} {\bibfnamefont {T.}~\bibnamefont {Garc\'{\i}a}}, \bibinfo {author}
  {\bibfnamefont {N.}~\bibnamefont {Guerrero}}, \bibinfo {author}
  {\bibfnamefont {A.~A.}\ \bibnamefont {Abbott}}, \bibinfo {author}
  {\bibfnamefont {M.}~\bibnamefont {Ara\'ujo}}, \bibinfo {author}
  {\bibfnamefont {C.}~\bibnamefont {Branciard}}, \bibinfo {author}
  {\bibfnamefont {E.~S.}\ \bibnamefont {G\'omez}}, \bibinfo {author}
  {\bibfnamefont {S.~P.}\ \bibnamefont {Walborn}}, \bibinfo {author}
  {\bibfnamefont {L.}~\bibnamefont {Aolita}},\ and\ \bibinfo {author}
  {\bibfnamefont {G.}~\bibnamefont {Lima}},\ }\href
  {https://doi.org/10.1103/PRXQuantum.2.010320} {\bibfield  {journal} {\bibinfo
   {journal} {PRX Quantum}\ }\textbf {\bibinfo {volume} {2}},\ \bibinfo {pages}
  {010320} (\bibinfo {year} {2021})}\BibitemShut {NoStop}%
\bibitem [{\citenamefont {Mart{\'i}nez}\ \emph {et~al.}(2022)\citenamefont
  {Mart{\'i}nez}, \citenamefont {G{\'o}mez}, \citenamefont {Cari{\~n}e},
  \citenamefont {Pereira}, \citenamefont {Delgado}, \citenamefont {Walborn},
  \citenamefont {Tavakoli},\ and\ \citenamefont {Lima}}]{martinez22}%
  \BibitemOpen
  \bibfield  {author} {\bibinfo {author} {\bibfnamefont {D.}~\bibnamefont
  {Mart{\'i}nez}}, \bibinfo {author} {\bibfnamefont {E.~S.}\ \bibnamefont
  {G{\'o}mez}}, \bibinfo {author} {\bibfnamefont {J.}~\bibnamefont
  {Cari{\~n}e}}, \bibinfo {author} {\bibfnamefont {L.}~\bibnamefont {Pereira}},
  \bibinfo {author} {\bibfnamefont {A.}~\bibnamefont {Delgado}}, \bibinfo
  {author} {\bibfnamefont {S.~P.}\ \bibnamefont {Walborn}}, \bibinfo {author}
  {\bibfnamefont {A.}~\bibnamefont {Tavakoli}},\ and\ \bibinfo {author}
  {\bibfnamefont {G.}~\bibnamefont {Lima}},\ }\bibfield  {journal} {\bibinfo
  {journal} {Nat. Phys.}\ }\href {https://doi.org/10.1038/s41567-022-01845-z}
  {10.1038/s41567-022-01845-z} (\bibinfo {year} {2022})\BibitemShut {NoStop}%
\bibitem [{\citenamefont {Lio}\ \emph {et~al.}(2020)\citenamefont {Lio},
  \citenamefont {Oxenlowe}, \citenamefont {Bacco}, \citenamefont {Cozzolino},
  \citenamefont {Biagi}, \citenamefont {Arge}, \citenamefont {Larsen},
  \citenamefont {Rottwitt}, \citenamefont {Ding},\ and\ \citenamefont
  {Zavatta}}]{lio20}%
  \BibitemOpen
  \bibfield  {author} {\bibinfo {author} {\bibfnamefont {B.~D.}\ \bibnamefont
  {Lio}}, \bibinfo {author} {\bibfnamefont {L.~K.}\ \bibnamefont {Oxenlowe}},
  \bibinfo {author} {\bibfnamefont {D.}~\bibnamefont {Bacco}}, \bibinfo
  {author} {\bibfnamefont {D.}~\bibnamefont {Cozzolino}}, \bibinfo {author}
  {\bibfnamefont {N.}~\bibnamefont {Biagi}}, \bibinfo {author} {\bibfnamefont
  {T.~N.}\ \bibnamefont {Arge}}, \bibinfo {author} {\bibfnamefont
  {E.}~\bibnamefont {Larsen}}, \bibinfo {author} {\bibfnamefont
  {K.}~\bibnamefont {Rottwitt}}, \bibinfo {author} {\bibfnamefont
  {Y.}~\bibnamefont {Ding}},\ and\ \bibinfo {author} {\bibfnamefont
  {A.}~\bibnamefont {Zavatta}},\ }\href@noop {} {\bibfield  {journal} {\bibinfo
   {journal} {IEEE J. Sel. Top. Quantum Electron.}\ }\textbf {\bibinfo {volume}
  {26}},\ \bibinfo {pages} {6400108} (\bibinfo {year} {2020})}\BibitemShut
  {NoStop}%
\bibitem [{\citenamefont {G\'omez}\ \emph {et~al.}(2021)\citenamefont
  {G\'omez}, \citenamefont {G\'omez}, \citenamefont {Machuca}, \citenamefont
  {Cabello}, \citenamefont {P\'adua}, \citenamefont {Walborn},\ and\
  \citenamefont {Lima}}]{gomez21}%
  \BibitemOpen
  \bibfield  {author} {\bibinfo {author} {\bibfnamefont {E.}~\bibnamefont
  {G\'omez}}, \bibinfo {author} {\bibfnamefont {S.}~\bibnamefont {G\'omez}},
  \bibinfo {author} {\bibfnamefont {I.}~\bibnamefont {Machuca}}, \bibinfo
  {author} {\bibfnamefont {A.}~\bibnamefont {Cabello}}, \bibinfo {author}
  {\bibfnamefont {S.}~\bibnamefont {P\'adua}}, \bibinfo {author} {\bibfnamefont
  {S.}~\bibnamefont {Walborn}},\ and\ \bibinfo {author} {\bibfnamefont
  {G.}~\bibnamefont {Lima}},\ }\href
  {https://doi.org/10.1103/PhysRevApplied.15.034024} {\bibfield  {journal}
  {\bibinfo  {journal} {Phys. Rev. Applied}\ }\textbf {\bibinfo {volume}
  {15}},\ \bibinfo {pages} {034024} (\bibinfo {year} {2021})}\BibitemShut
  {NoStop}%
\bibitem [{\citenamefont {Wu}\ \emph {et~al.}(2025)\citenamefont {Wu},
  \citenamefont {Wang}, \citenamefont {Zhang}, \citenamefont {Yu},
  \citenamefont {Li}, \citenamefont {Guo}, \citenamefont {Wang}, \citenamefont
  {Zhang}, \citenamefont {Wang}, \citenamefont {Chen},\ and\ \citenamefont
  {Chen}}]{Wu2025}%
  \BibitemOpen
  \bibfield  {author} {\bibinfo {author} {\bibfnamefont {Q.}~\bibnamefont
  {Wu}}, \bibinfo {author} {\bibfnamefont {C.}~\bibnamefont {Wang}}, \bibinfo
  {author} {\bibfnamefont {H.}~\bibnamefont {Zhang}}, \bibinfo {author}
  {\bibfnamefont {Z.}~\bibnamefont {Yu}}, \bibinfo {author} {\bibfnamefont
  {Y.}~\bibnamefont {Li}}, \bibinfo {author} {\bibfnamefont {Y.}~\bibnamefont
  {Guo}}, \bibinfo {author} {\bibfnamefont {H.}~\bibnamefont {Wang}}, \bibinfo
  {author} {\bibfnamefont {D.}~\bibnamefont {Zhang}}, \bibinfo {author}
  {\bibfnamefont {T.}~\bibnamefont {Wang}}, \bibinfo {author} {\bibfnamefont
  {Z.}~\bibnamefont {Chen}},\ and\ \bibinfo {author} {\bibfnamefont
  {X.}~\bibnamefont {Chen}},\ }\href
  {https://doi.org/10.1038/s41377-025-01982-z} {\bibfield  {journal} {\bibinfo
  {journal} {Light: Science \& Applications}\ }\textbf {\bibinfo {volume}
  {13}},\ \bibinfo {pages} {19} (\bibinfo {year} {2025})}\BibitemShut {NoStop}%
\bibitem [{\citenamefont {Xiong}\ \emph {et~al.}(2018)\citenamefont {Xiong},
  \citenamefont {Hsu}, \citenamefont {Bromberg}, \citenamefont {Oz},
  \citenamefont {Lerman}, \citenamefont {Chang},\ and\ \citenamefont
  {Cao}}]{Xiong2018}%
  \BibitemOpen
  \bibfield  {author} {\bibinfo {author} {\bibfnamefont {W.}~\bibnamefont
  {Xiong}}, \bibinfo {author} {\bibfnamefont {C.~W.}\ \bibnamefont {Hsu}},
  \bibinfo {author} {\bibfnamefont {Y.}~\bibnamefont {Bromberg}}, \bibinfo
  {author} {\bibfnamefont {H.~M.}\ \bibnamefont {Oz}}, \bibinfo {author}
  {\bibfnamefont {N.~M.}\ \bibnamefont {Lerman}}, \bibinfo {author}
  {\bibfnamefont {S.~J.~B.}\ \bibnamefont {Chang}},\ and\ \bibinfo {author}
  {\bibfnamefont {H.}~\bibnamefont {Cao}},\ }\href
  {https://doi.org/10.1038/s41377-018-0047-4} {\bibfield  {journal} {\bibinfo
  {journal} {Light: Science \& Applications}\ }\textbf {\bibinfo {volume}
  {7}},\ \bibinfo {pages} {47} (\bibinfo {year} {2018})}\BibitemShut {NoStop}%
\bibitem [{\citenamefont {dos Santos}\ \emph {et~al.}(2026)\citenamefont {dos
  Santos}, \citenamefont {Sawada}, \citenamefont {Villalba}, \citenamefont
  {Jara}, \citenamefont {Guerrero}, \citenamefont {Melo}, \citenamefont
  {Magiotto}, \citenamefont {Martínez}, \citenamefont {Xavier}, \citenamefont
  {Cariñe}, \citenamefont {Saavedra}, \citenamefont {Gómez}, \citenamefont
  {Walborn},\ and\ \citenamefont
  {Lima}}]{santos2026highdimensionalquantumkeydistribution}%
  \BibitemOpen
  \bibfield  {author} {\bibinfo {author} {\bibfnamefont {G.~H.}\ \bibnamefont
  {dos Santos}}, \bibinfo {author} {\bibfnamefont {K.~B.}\ \bibnamefont
  {Sawada}}, \bibinfo {author} {\bibfnamefont {N.}~\bibnamefont {Villalba}},
  \bibinfo {author} {\bibfnamefont {C.}~\bibnamefont {Jara}}, \bibinfo {author}
  {\bibfnamefont {N.}~\bibnamefont {Guerrero}}, \bibinfo {author}
  {\bibfnamefont {C.}~\bibnamefont {Melo}}, \bibinfo {author} {\bibfnamefont
  {M.~H.}\ \bibnamefont {Magiotto}}, \bibinfo {author} {\bibfnamefont
  {D.}~\bibnamefont {Martínez}}, \bibinfo {author} {\bibfnamefont {G.~B.}\
  \bibnamefont {Xavier}}, \bibinfo {author} {\bibfnamefont {J.}~\bibnamefont
  {Cariñe}}, \bibinfo {author} {\bibfnamefont {G.}~\bibnamefont {Saavedra}},
  \bibinfo {author} {\bibfnamefont {E.~S.}\ \bibnamefont {Gómez}}, \bibinfo
  {author} {\bibfnamefont {S.~P.}\ \bibnamefont {Walborn}},\ and\ \bibinfo
  {author} {\bibfnamefont {G.}~\bibnamefont {Lima}},\ }\href
  {https://arxiv.org/abs/2606.04211} {\bibinfo {title} {High-dimensional
  quantum key distribution via full core-mode encoding over deployed multicore
  fibers}} (\bibinfo {year} {2026}),\ \Eprint
  {https://arxiv.org/abs/2606.04211} {arXiv:2606.04211 [quant-ph]} \BibitemShut
  {NoStop}%
\bibitem [{\citenamefont {Cari{\~n}e}\ \emph {et~al.}(2020)\citenamefont
  {Cari{\~n}e}, \citenamefont {Ca{\~n}as}, \citenamefont {Skrzypczyk},
  \citenamefont {\v{S}upi\'{c}}, \citenamefont {Guerrero}, \citenamefont
  {Garcia}, \citenamefont {Pereira}, \citenamefont {Prosser}, \citenamefont
  {Xavier}, \citenamefont {Delgado}, \citenamefont {Walborn}, \citenamefont
  {Cavalcanti},\ and\ \citenamefont {Lima}}]{carine20}%
  \BibitemOpen
  \bibfield  {author} {\bibinfo {author} {\bibfnamefont {J.}~\bibnamefont
  {Cari{\~n}e}}, \bibinfo {author} {\bibfnamefont {G.}~\bibnamefont
  {Ca{\~n}as}}, \bibinfo {author} {\bibfnamefont {P.}~\bibnamefont
  {Skrzypczyk}}, \bibinfo {author} {\bibfnamefont {I.}~\bibnamefont
  {\v{S}upi\'{c}}}, \bibinfo {author} {\bibfnamefont {N.}~\bibnamefont
  {Guerrero}}, \bibinfo {author} {\bibfnamefont {T.}~\bibnamefont {Garcia}},
  \bibinfo {author} {\bibfnamefont {L.}~\bibnamefont {Pereira}}, \bibinfo
  {author} {\bibfnamefont {M.~A.~S.}\ \bibnamefont {Prosser}}, \bibinfo
  {author} {\bibfnamefont {G.~B.}\ \bibnamefont {Xavier}}, \bibinfo {author}
  {\bibfnamefont {A.}~\bibnamefont {Delgado}}, \bibinfo {author} {\bibfnamefont
  {S.~P.}\ \bibnamefont {Walborn}}, \bibinfo {author} {\bibfnamefont
  {D.}~\bibnamefont {Cavalcanti}},\ and\ \bibinfo {author} {\bibfnamefont
  {G.}~\bibnamefont {Lima}},\ }\href {https://doi.org/10.1364/OPTICA.388912}
  {\bibfield  {journal} {\bibinfo  {journal} {Optica}\ }\textbf {\bibinfo
  {volume} {7}},\ \bibinfo {pages} {542} (\bibinfo {year} {2020})}\BibitemShut
  {NoStop}%
\bibitem [{\citenamefont {Silva-Lopez}\ \emph {et~al.}(2004)\citenamefont
  {Silva-Lopez}, \citenamefont {Li}, \citenamefont {MacPherson}, \citenamefont
  {Moore}, \citenamefont {Barton}, \citenamefont {Jones}, \citenamefont {Zhao},
  \citenamefont {Zhang},\ and\ \citenamefont {Bennion}}]{silva-lopez2004}%
  \BibitemOpen
  \bibfield  {author} {\bibinfo {author} {\bibfnamefont {M.}~\bibnamefont
  {Silva-Lopez}}, \bibinfo {author} {\bibfnamefont {C.}~\bibnamefont {Li}},
  \bibinfo {author} {\bibfnamefont {W.~N.}\ \bibnamefont {MacPherson}},
  \bibinfo {author} {\bibfnamefont {A.~J.}\ \bibnamefont {Moore}}, \bibinfo
  {author} {\bibfnamefont {J.~S.}\ \bibnamefont {Barton}}, \bibinfo {author}
  {\bibfnamefont {J.~D.~C.}\ \bibnamefont {Jones}}, \bibinfo {author}
  {\bibfnamefont {D.}~\bibnamefont {Zhao}}, \bibinfo {author} {\bibfnamefont
  {L.}~\bibnamefont {Zhang}},\ and\ \bibinfo {author} {\bibfnamefont
  {I.}~\bibnamefont {Bennion}},\ }\href {https://doi.org/10.1364/OL.29.002225}
  {\bibfield  {journal} {\bibinfo  {journal} {Optics Letters}\ }\textbf
  {\bibinfo {volume} {29}},\ \bibinfo {pages} {2225} (\bibinfo {year}
  {2004})}\BibitemShut {NoStop}%
\bibitem [{\citenamefont {Silva-L{\'o}pez}\ \emph {et~al.}(2005)\citenamefont
  {Silva-L{\'o}pez}, \citenamefont {MacPherson}, \citenamefont {Li},
  \citenamefont {Moore}, \citenamefont {Barton}, \citenamefont {Jones},
  \citenamefont {Zhao}, \citenamefont {Zhang},\ and\ \citenamefont
  {Bennion}}]{silva-lopez2005}%
  \BibitemOpen
  \bibfield  {author} {\bibinfo {author} {\bibfnamefont {M.}~\bibnamefont
  {Silva-L{\'o}pez}}, \bibinfo {author} {\bibfnamefont {W.~N.}\ \bibnamefont
  {MacPherson}}, \bibinfo {author} {\bibfnamefont {C.}~\bibnamefont {Li}},
  \bibinfo {author} {\bibfnamefont {A.~J.}\ \bibnamefont {Moore}}, \bibinfo
  {author} {\bibfnamefont {J.~S.}\ \bibnamefont {Barton}}, \bibinfo {author}
  {\bibfnamefont {J.~D.~C.}\ \bibnamefont {Jones}}, \bibinfo {author}
  {\bibfnamefont {D.}~\bibnamefont {Zhao}}, \bibinfo {author} {\bibfnamefont
  {L.}~\bibnamefont {Zhang}},\ and\ \bibinfo {author} {\bibfnamefont
  {I.}~\bibnamefont {Bennion}},\ }\href {https://doi.org/10.1364/AO.44.006890}
  {\bibfield  {journal} {\bibinfo  {journal} {Applied Optics}\ }\textbf
  {\bibinfo {volume} {44}},\ \bibinfo {pages} {6890} (\bibinfo {year}
  {2005})}\BibitemShut {NoStop}%
\bibitem [{\citenamefont {Flockhart}\ \emph {et~al.}(2003)\citenamefont
  {Flockhart}, \citenamefont {MacPherson}, \citenamefont {Barton},
  \citenamefont {Jones}, \citenamefont {Zhang},\ and\ \citenamefont
  {Bennion}}]{flockhart03}%
  \BibitemOpen
  \bibfield  {author} {\bibinfo {author} {\bibfnamefont {G.~M.~H.}\
  \bibnamefont {Flockhart}}, \bibinfo {author} {\bibfnamefont {W.~N.}\
  \bibnamefont {MacPherson}}, \bibinfo {author} {\bibfnamefont {J.~S.}\
  \bibnamefont {Barton}}, \bibinfo {author} {\bibfnamefont {J.~D.~C.}\
  \bibnamefont {Jones}}, \bibinfo {author} {\bibfnamefont {L.}~\bibnamefont
  {Zhang}},\ and\ \bibinfo {author} {\bibfnamefont {I.}~\bibnamefont
  {Bennion}},\ }\href {https://doi.org/10.1364/OL.28.000387} {\bibfield
  {journal} {\bibinfo  {journal} {Opt. Lett.}\ }\textbf {\bibinfo {volume}
  {28}},\ \bibinfo {pages} {387} (\bibinfo {year} {2003})}\BibitemShut
  {NoStop}%
\bibitem [{\citenamefont {Villatoro}\ \emph {et~al.}(2016)\citenamefont
  {Villatoro}, \citenamefont {Van~Newkirk}, \citenamefont {Antonio-Lopez},
  \citenamefont {Zubia}, \citenamefont {Sch\"{u}lzgen},\ and\ \citenamefont
  {Amezcua-Correa}}]{villatoro16}%
  \BibitemOpen
  \bibfield  {author} {\bibinfo {author} {\bibfnamefont {J.}~\bibnamefont
  {Villatoro}}, \bibinfo {author} {\bibfnamefont {A.}~\bibnamefont
  {Van~Newkirk}}, \bibinfo {author} {\bibfnamefont {E.}~\bibnamefont
  {Antonio-Lopez}}, \bibinfo {author} {\bibfnamefont {J.}~\bibnamefont
  {Zubia}}, \bibinfo {author} {\bibfnamefont {A.}~\bibnamefont
  {Sch\"{u}lzgen}},\ and\ \bibinfo {author} {\bibfnamefont {R.}~\bibnamefont
  {Amezcua-Correa}},\ }\href {https://doi.org/10.1364/OL.41.000832} {\bibfield
  {journal} {\bibinfo  {journal} {Opt. Lett.}\ }\textbf {\bibinfo {volume}
  {41}},\ \bibinfo {pages} {832} (\bibinfo {year} {2016})}\BibitemShut
  {NoStop}%
\bibitem [{\citenamefont {Moore}\ and\ \citenamefont {Rogge}(2012)}]{moore12}%
  \BibitemOpen
  \bibfield  {author} {\bibinfo {author} {\bibfnamefont {J.~P.}\ \bibnamefont
  {Moore}}\ and\ \bibinfo {author} {\bibfnamefont {M.~D.}\ \bibnamefont
  {Rogge}},\ }\href {https://doi.org/10.1364/OE.20.002967} {\bibfield
  {journal} {\bibinfo  {journal} {Opt. Express}\ }\textbf {\bibinfo {volume}
  {20}},\ \bibinfo {pages} {2967} (\bibinfo {year} {2012})}\BibitemShut
  {NoStop}%
\bibitem [{\citenamefont {Zhao}\ \emph {et~al.}(2022)\citenamefont {Zhao},
  \citenamefont {Dang},\ and\ \citenamefont {Tang}}]{zhao2022advances}%
  \BibitemOpen
  \bibfield  {author} {\bibinfo {author} {\bibfnamefont {Z.}~\bibnamefont
  {Zhao}}, \bibinfo {author} {\bibfnamefont {Y.}~\bibnamefont {Dang}},\ and\
  \bibinfo {author} {\bibfnamefont {M.}~\bibnamefont {Tang}},\ }\href
  {https://doi.org/10.3390/photonics9060381} {\bibfield  {journal} {\bibinfo
  {journal} {Photonics}\ }\textbf {\bibinfo {volume} {9}},\ \bibinfo {pages}
  {381} (\bibinfo {year} {2022})}\BibitemShut {NoStop}%
\bibitem [{\citenamefont {Ocampo}\ and\ \citenamefont
  {Saitoh}(2025)}]{ocampo25}%
  \BibitemOpen
  \bibfield  {author} {\bibinfo {author} {\bibfnamefont {G.}~\bibnamefont
  {Ocampo}}\ and\ \bibinfo {author} {\bibfnamefont {K.}~\bibnamefont
  {Saitoh}},\ }\href {https://doi.org/10.1364/OE.558028} {\bibfield  {journal}
  {\bibinfo  {journal} {Opt. Express}\ }\textbf {\bibinfo {volume} {33}},\
  \bibinfo {pages} {16052} (\bibinfo {year} {2025})}\BibitemShut {NoStop}%
\bibitem [{\citenamefont {Diamandi}\ \emph {et~al.}(2017)\citenamefont
  {Diamandi}, \citenamefont {London},\ and\ \citenamefont
  {Zadok}}]{diamandi17}%
  \BibitemOpen
  \bibfield  {author} {\bibinfo {author} {\bibfnamefont {H.~H.}\ \bibnamefont
  {Diamandi}}, \bibinfo {author} {\bibfnamefont {Y.}~\bibnamefont {London}},\
  and\ \bibinfo {author} {\bibfnamefont {A.}~\bibnamefont {Zadok}},\ }\href
  {https://doi.org/10.1364/OPTICA.4.000289} {\bibfield  {journal} {\bibinfo
  {journal} {Optica}\ }\textbf {\bibinfo {volume} {4}},\ \bibinfo {pages} {289}
  (\bibinfo {year} {2017})}\BibitemShut {NoStop}%
\bibitem [{\citenamefont {{Thorlabs,
  Inc.}}(2014)}]{Thorlabs_PolarizationController_2014}%
  \BibitemOpen
  \bibfield  {author} {\bibinfo {author} {\bibnamefont {{Thorlabs, Inc.}}},\
  }\href@noop {} {\bibinfo {title} {In‑line fiber optic polarization
  controllers}},\ \bibinfo {howpublished}
  {\url{https://www.thorlabs.com/newgrouppage9.cfm?objectgroup_id=2161}}
  (\bibinfo {year} {2014}),\ \bibinfo {note} {last edited March 5, 2014;
  accessed June 8, 2025}\BibitemShut {NoStop}%
\bibitem [{\citenamefont {OptoSigma}(2024)}]{OptoSigma_PC_Web}%
  \BibitemOpen
  \bibfield  {author} {\bibinfo {author} {\bibnamefont {OptoSigma}},\ }\href
  {https://sea.optosigma.com/en_sg/ipc-250.html} {\bibinfo {title} {In-line
  fiber polarization controllers / ipc-250}} (\bibinfo {year}
  {2024})\BibitemShut {NoStop}%
\bibitem [{\citenamefont {{Fibercore}}(2026)}]{fibercore_datasheet}%
  \BibitemOpen
  \bibfield  {author} {\bibinfo {author} {\bibnamefont {{Fibercore}}},\
  }\href@noop {} {\bibinfo {title} {{SM-4C1500(8.0/125)/001 Multicore Fiber
  Product Page}}},\ \bibinfo {howpublished}
  {\url{https://fibercore.humaneticsgroup.com/products/multicore-fiber/multicore-fiber/sm-4c150080125001}}
  (\bibinfo {year} {2026}),\ \bibinfo {note} {accessed: 2026-05-11}\BibitemShut
  {NoStop}%
\bibitem [{\citenamefont {Johnson}(1979)}]{johnson1979inline}%
  \BibitemOpen
  \bibfield  {author} {\bibinfo {author} {\bibfnamefont {M.}~\bibnamefont
  {Johnson}},\ }\href {https://doi.org/10.1364/AO.18.001288} {\bibfield
  {journal} {\bibinfo  {journal} {Applied Optics}\ }\textbf {\bibinfo {volume}
  {18}},\ \bibinfo {pages} {1288} (\bibinfo {year} {1979})}\BibitemShut
  {NoStop}%
\bibitem [{\citenamefont {Smith}(1980)}]{smith1980pressure}%
  \BibitemOpen
  \bibfield  {author} {\bibinfo {author} {\bibfnamefont {A.~M.}\ \bibnamefont
  {Smith}},\ }\href {https://doi.org/10.1049/el:19800534} {\bibfield  {journal}
  {\bibinfo  {journal} {Electronics Letters}\ }\textbf {\bibinfo {volume}
  {16}},\ \bibinfo {pages} {773} (\bibinfo {year} {1980})}\BibitemShut
  {NoStop}%
\bibitem [{\citenamefont {Frocht}(1941)}]{Frocht1941Photoelasticity}%
  \BibitemOpen
  \bibfield  {author} {\bibinfo {author} {\bibfnamefont {M.~M.}\ \bibnamefont
  {Frocht}},\ }\href@noop {} {\emph {\bibinfo {title} {Photoelasticity}}},\
  Vol.~\bibinfo {volume} {1}\ (\bibinfo  {publisher} {John Wiley \& Sons},\
  \bibinfo {address} {New York},\ \bibinfo {year} {1941})\BibitemShut {NoStop}%
\bibitem [{\citenamefont {Gianino}\ and\ \citenamefont
  {Bendow}(1981)}]{gianino81}%
  \BibitemOpen
  \bibfield  {author} {\bibinfo {author} {\bibfnamefont {P.~D.}\ \bibnamefont
  {Gianino}}\ and\ \bibinfo {author} {\bibfnamefont {B.}~\bibnamefont
  {Bendow}},\ }\href {https://doi.org/10.1364/AO.20.000430} {\bibfield
  {journal} {\bibinfo  {journal} {Appl. Opt.}\ }\textbf {\bibinfo {volume}
  {20}},\ \bibinfo {pages} {430} (\bibinfo {year} {1981})}\BibitemShut
  {NoStop}%
\bibitem [{\citenamefont {Sadd}(2005)}]{sadd2005elasticity}%
  \BibitemOpen
  \bibfield  {author} {\bibinfo {author} {\bibfnamefont {M.~H.}\ \bibnamefont
  {Sadd}},\ }\href@noop {} {\emph {\bibinfo {title} {Elasticity: Theory,
  Applications, and Numerics}}},\ \bibinfo {edition} {1st}\ ed.\ (\bibinfo
  {publisher} {Elsevier Academic Press},\ \bibinfo {year} {2005})\BibitemShut
  {NoStop}%
\end{thebibliography}
\end{document}